\RequirePackage{fix-cm}

\documentclass[authoryear,11pt]{elsarticle}

\usepackage{amsmath}
\usepackage{amsfonts}
\usepackage{graphicx}
\usepackage{dcolumn}
\usepackage{bm}
\usepackage{amssymb}
\usepackage{latexsym}
\usepackage{color}
\usepackage{url}
\usepackage[round]{natbib}
\usepackage{enumerate}

\newcommand{\pbrac}[1]{\left( #1 \right)}
\newcommand{\tbrac}[1]{\left[ #1 \right]}
\newcommand{\cbrac}[1]{\left\{ #1 \right\}}

\numberwithin{equation}{section}

\begin{document}

\begin{frontmatter}

\title{An Analytical Approach to (Meta)Relational Models Theory\\and its Application to Triple Bottom Line (Profit, People, Planet)\\ \textit{Towards Social Relations Portfolio Management}}

\author{Arsham Farzinnia\corref{cor1}}
\ead{farzinni@msu.edu}
\cortext[cor1]{ORCID: 0000-0002-2883-8601 (corresponding author)}

\author{Corine Boon\corref{cor2}}
\ead{c.t.boon@uva.nl}
\cortext[cor2]{ORCID: 0000-0001-7245-4923}

\address{Amsterdam Business School, University of Amsterdam, P.O. Box 19268, 1000 GG Amsterdam, The Netherlands}

\date{\today}

\begin{abstract}

Investigating the optimal nature of social interactions among actors (e.g., people or firms), who seek to achieve certain mutually-agreed objectives, has been the subject of extensive academic research. Using the relational models theory (describing all social interactions as combinations of four basic sociality ingredients: Communal Sharing, Authority Ranking, Equality Matching, and Market Pricing), the common approach revolves around \textit{qualitative} arguments for determining sociality configurations most effective in realizing specific purposes, at times supplemented by empirical data. In the current treatment, we formulate this question as a mathematical optimization problem, in order to \textit{quantitatively} derive the most suitable combination of sociality forms for dyadic actors, which optimizes their mutually-agreed objective. For this purpose, we develop an analytical framework of the (meta)relational models theory, and demonstrate that combining the four sociality forms to define a specific meaningful social situation inevitably prompts an inherent tension among them, codified by a single elementary and universal metarelation. In analogy with financial portfolio management, we subsequently introduce the concept of \textit{Social Relations Portfolio} (SRP) management, and propose a generalizable methodology capable of quantitatively identifying the \textit{efficient} SRP, which, in turn, enables effective stakeholder and change management initiatives. As an important illustration, the methodology is applied to the Triple Bottom Line (Profit, People, Planet) paradigm to derive its efficient SRP. This serves as a guide to practitioners for precisely measuring, monitoring, reporting and steering stakeholder and change management efforts concerning Corporate Social Responsibility (CSR) and Environmental, Social and Governance (ESG) within and~/ or across organizations.

\end{abstract}


\end{frontmatter}

\newpage

\setcounter{tocdepth}{3}
\tableofcontents

\newpage


\section{Introduction}\label{Intr}

Systematic understanding of social relations among humans, and their (generally complex) nature, has been the subject of fascinating scientific research for many decades. The pioneering theoretical attempts in this field primarily comprised of a myriad of independent and unconnected situational paradigms, pertaining to distinct groups of social occurrences. Building upon and complementing the earlier findings (e.g., by \cite{Deutsch:1975}), the \textit{relational models theory} \citep{Fiske:1991, Fiske:1992} proposes a unified framework and a theoretical foundation for understanding, describing and simplifying the complex nature of social relations, with influential applications in various fields of research (c.f., Section~\ref{LitRev}). In particular,

\begin{quote}
``The [relational models] theory postulates that people in all cultures use just four relational models to generate most kinds of social interaction, evaluation, and affect. People construct complex and varied social forms using combinations of these models implemented according to diverse cultural rules.'' \citep[p.689]{Fiske:1992}
\end{quote}

According to this seminal work, the socially meaningful interactions are categorized into four basic sociality schemata (or relational models) of \textit{Communal Sharing}, \textit{Authority Ranking}, \textit{Equality Matching} and \textit{Market Pricing}, the various combinations of which entail the observed rich and complex human relationships \citep{Sheppard:1996, Fiske:2005}. In other words, any meaningful social interaction is proposed to be reducible to a combination of the aforementioned four basic relational models. These models are understood to operate autonomously and independently from one another as distinct structures. 

The combinatorics of relational models were further studied within the \textit{metarelational models} framework \citep{Fiske:2012}, which governs configurations of the four basic sociality forms, specifying how relationships are (or are not) supposed to be combined. The paramount importance of these combinations in regulating social systems was recognized as,

\begin{quote}
``We cannot understand individual psychology, dyadic social relationships, or group psychology unless we understand the combinatorics of social relationships\dots These combinatorial, syntactic configurations of social relationships are important features of cultures.'' \citep[p.2]{Fiske:2012}
\end{quote}

Considering relational models as basic ingredients or `syntactical' for conjoining relationships, metarelational models are thus specifically concerned with the combinatorics of the four sociality forms that inform, entail and~/ or preclude one another. The framework considers various possible configurations of social interactions between two actors, as well as (higher order) relations among several parties, and explores their range of applicability and potential consequences. 

Be that as it may, several open questions concerning the metarelational models still remain to be examined \citep{Fiske:2012}. Most importantly, there is a lack of knowledge on the constraints on how specific relational models combine, and whether \textit{universal} metarelational models exist with elementary innate structures and combinatorial proclivities. Furthermore, as mentioned earlier, the relational models theory has successfully been utilized in a multitude of fields of theoretical and applied research involving human relations and~/ or stakeholder management, where the most suitable model, or combination of models (i.e., metarelational model), for fulfilling a particular predefined purpose is investigated.\footnote{For instance, studies have focused on the best relational model (mixture) among stakeholders which would enhance or optimize helping behavior (e.g., \cite{Mossholder:2011}), supply chain management (e.g., \cite{Lejeune:2005} and \cite{Blomme:2014}), or joint value creation (e.g., \cite{Bridoux:2016}) --- see Section~\ref{LitRev} for a review.} Thus far, however, within the literature the obtained results and drawn conclusions are largely based on \textit{qualitative} arguments, at times supported by gathered empirical (statistical) data. One notes that in this type of application, one is essentially faced with an optimization problem, which is most effectively tackled using \textit{quantitative} methods. Consequently, an analytical understanding of the (meta)relational models theory and its applications, at a theoretical level, is currently lacking, in spite of its evident importance and necessity.\footnote{It should be mentioned that the current state of mathematical modeling efforts within social sciences appears to be severely underrepresented compared with other fields (such as natural sciences or finance), and hence in dire need of adequate attention.}

In the current treatment, we propose concrete solutions to address the two abovementioned gaps, by specifically articulating and answering the following main Research Question:
\begin{quote}
\textbf{RQ}: \emph{What is the optimal relative configuration of the four relational models within a meaningful social interaction between two actors, who intend to achieve a certain mutually-agreed objective?}
\end{quote}
To this end, we approach the metarelational models theory as a mathematical optimization problem, aiming at constructing a generalizable framework with a broad range of applicability. The overall contribution of our interdisciplinary work is, hence, threefold and can be summarized as follows:

First, we develop an analytical framework for the relational models theory, quantifying its parameters as well as their metarelational interconnection. The mathematical model enables to demonstrate, among others, that the original assumption regarding autonomy and independence of the four relational models cannot hold once they are combined to define a social situation; specifically, within the metarelational context, an inherent \textit{universal} tension among all four models is identified (e.g., between the ``relational'' vs.~``transactional'' views, or the ``prosocial'' vs.~``self-oriented'' relations) and concretized by a single metarelation. In other words, we analytically show that combining the relational models to describe any particular social interaction inevitably prompts internal preclusion (i.e., compromises or tradeoffs) among them, according to one universal elementary metarelation. This addresses the first gap mentioned above.

Second, using the developed model, we propose a generalizable methodology capable of determining the optimal configuration of four relational models within a meaningful social interaction, given predefined objectives. In this manner, we bridge the second aforementioned gap, by solving the optimization problem ubiquitously encountered within the literature. A corresponding systematic procedure is devised and detailed for conducting the pertaining optimization operations. Moreover, in analogy with financial asset~/ portfolio management in economics, this methodology practically enables \textit{Social Relations Portfolio} management, by identifying the corresponding \textit{efficient} Social Relations Portfolio for actors who intend to optimally achieve certain interaction objective(s). The introduced methodology is general in nature, and can be applied to optimize any arbitrary social interaction goal involving metarelational models; it is, therefore, particularly relevant in measuring, optimizing, and reporting organizational culture, inter-organizational relations, stakeholder management and change management.

To illustrate this in practice, we apply the proposed method to the Triple Bottom Line (TBL) paradigm \citep{Elkington:1998, Elkington:2018, Slaper:2011}.\footnote{An application of the quantitative framework of the relational models theory to optimizing joint value creation \citep{Bridoux:2016} and the United Nations Sustainable Development Goals is examined by \cite{Farzinnia:2019}.} The TBL paradigm (along with various other `Green' facets) concretizes the sustainable business model by simultaneously taking into account the three pillars of \textit{Profit}, \textit{People}, and \textit{Planet} (3Ps), and is increasingly adopted by organizations as part of their Corporate Social Responsibility (CSR) and Environmental, Social and Governance (ESG) endeavors. In accordance, we determine the optimal relative combination of relational models to be applied within an organization (e.g., a firm or a collection of firms) which would result in maximizing TBL and, thus, socially- and environmentally-responsible business conduct. That is to say, we aim to identify TBL's efficient Social Relations Portfolio, which would, in turn, reflect the desired organizational culture and~/ or inter-organizational relations with respect to CSR~/ ESG efforts. Guided by the vast available research literature, we utilize productivity, social well-being, and longterm sustainability as (measurable) proxies for \textit{Profit}, \textit{People}, and \textit{Planet}, respectively, and construct the analytical behavior of each of these pillars as a function of the four relational models. These mathematical representations are then employed to derive the optimal value of TBL's Social Relations Portfolio. In this manner, a concrete method is presented and elaborated for practitioners to engage in TBL optimization, measuring and reporting \citep{Goel:2010}.

Finally, in an interdisciplinary fashion, we bring together the extensive (although largely unconnected) literature on pertaining subjects in economics, sociology, stakeholder management, organizational culture, and sustainability, and interconnect them under one unified, coherent umbrella framework. In particular, this is manifested by the construction of TBL pillars' behaviors as functions of the sociality forms in different regimes of their strengths, which requires careful integration of insights from several relevant disciplines. Furthermore, the presented mathematical analysis is primarily inspired by established techniques widely utilized in natural sciences (e.g., theoretical physics). As a consequence, the overall mindset, approach and rigor introduced in this interdisciplinary treatment not only pertains to the discussed (meta)relational models theory and its application to TBL optimization, but may additionally serve as guidance for comprehensive mathematical modeling endeavors in social~/ behavioral sciences, allowing for obtaining quantitative results, thereby opening up new avenues for precision measurement in these important fields.

The paper is organized as follows: in Section~\ref{LitRev}, the relational and metarelational models theories and their most important~/ relevant features are explicated. Additionally, a review of the literature regarding their application in various areas is provided. The section concludes by discussing the TBL paradigm, and its (more recent) applications and explorations within the literature. Section~\ref{Funrel} introduces the method; specifically, the analytical approach to (meta)relational models theory, as well as the concept of Social Relations Portfolio management. A generalizable methodology for dealing with the corresponding optimization problem in a procedural manner is presented. To illustrate the broad range of applicability of the developed methodology, it is subsequently applied to TBL in Section~\ref{Applic} as a concrete case in point. Here, for each of the three pillars, the corresponding behavior as a function of sociality forms is derived, motivated by the available literature. Using mathematical representations of these profiles, the optimal relative configurations of sociality forms maximizing pillars and TBL (i.e., their efficient Social Relations Portfolios) are determined. The implications of obtained results, as well as several limitations of the current treatment, are further discussed in Section~\ref{Implic}, where potential avenues for future research are also outlined. Finally, we summarize and conclude the main contributions of this work in Section~\ref{Disc}. For reference and convenience, a summary of analytical approach to (meta)relational models theory, constructed mathematical profiles of TBL pillars as functions of the sociality forms, along with their corresponding conditions and illustrative values, is provided in Appendix.


\section{Literature Review}\label{LitRev}

According to the relational models theory \citep{Fiske:1991, Fiske:1992, Fiske:2004}, ``the motivation, planning, production, comprehension, coordination, and evaluation of human social life may be based largely on combinations of four psychological models''; namely, \textit{Communal Sharing} (CS), \textit{Authority Ranking} (AR), \textit{Equality Matching} (EM), and \textit{Market Pricing} (MP). These four sociality schemata, hence, constitute the elementary building blocks for constructing rich and complex meaningful social relations among humans \citep{Sheppard:1996, Fiske:2005}.\footnote{For a formal substantiation of why there may only be four relational models (along with the \textit{asocial} and \textit{null} interactions), see the interesting analyses by \cite{Favre:2013, Favre:2015}, as well as the discussion by \cite{Fiske:1992}.}

In CS, all group members are undifferentiated and equivalent to one another; there are no individual privileges or disadvantages, and no one member takes precedence over another \citep{Deutsch:1975, Bridoux:2016, Stofberg:2021}. Group members contribute what they can, in an altruistic manner, and take what they need, without an explicit bookkeeping mechanism or reciprocation commitment~/ expectation in place \citep{Gittell:2012}. The notion of `self' is largely fused with that of the collective and members strongly identify with their community (`extended self'), sharing a sense of unity to promote community's interests. Fairness of resource distribution is judged by the `need' of each member, and decision making rests on consensus. Group collectivism takes precedence over individualism and personal identity, and the model primarily satisfies members' need for belonging to a (social) collective. The socially meaningful relationships are, thus, defined by a nominal scale \citep{Fiske:1992}.

In contrast, AR is characterized by an explicit asymmetric linear hierarchy among elements of the group, which is accepted as fair, upheld and respected by them. In this model, everyone's rank can be compared to others', and their rank within the hierarchy also determines their identity. Fairness of resource distribution is determined by the respective `status' of members, and decision making proceeds through a chain of command from superior to subordinate. It is important to note that the superior--subordinate hierarchal relationship is perceived to rest on a legitimate and just social contract \citep{Tyler:2006}, in which the superiors' need for authority, power and dominance is voluntarily recognized by the subordinates, in exchange for satisfying their own need for leadership, protection, and patriarchal care, akin to the notion of \textit{noblesse oblige} \citep{Giessner:2010}. The meaningful social relations take the form of an ordinal scale of measure \citep{Fiske:1992}.

The socially meaningful relations based on EM are analogous to the familiar phrases ``I scratch your back, you scratch mine!'', \textit{quid pro quo}, or the `tit-for-tat' rule on an equal footing \citep{Deutsch:1975, Giessner:2010}, and are in line with the concept of social exchange \citep{Blau:1964} and balanced reciprocity \citep{Fiske:1991, Fehr:2002}. In other words, there is a prominent bookkeeping in place of interactions' imbalances, and reciprocity resides at the center of relations. Members ought to return the `favors' they owe in kind, as well as to expect in-kind punishment~/ retaliation for any `harm' they might afflict. Actors perceive one another as equal partners, and the model primarily addresses their need for equal voice and treatment. Fairness of resource distribution is assessed by the strict `equality' of everyone's rights and obligations, and decision making occurs through participants' equal say (`one person, one vote'). The operations under this model follow addition and subtraction rules to determine any imbalance, and, hence, correspond to interval measurements \citep{Fiske:1992}.

Finally, MP represents the form of sociality in which an understanding of multiplication, division, proportions, as well as distributive laws is required by group elements. In this model, personal identity and individualism are saliently present, often playing a central role in relationships \citep{Deutsch:1975}; thus, the model satisfies these needs. Social relations can be perceived as transactions; a proportionate common `price' or `exchange rate' may (symbolically) be agreed upon and assigned to the value of a mutual interaction (in a broad sense), allowing members to compare and organize their interactions with reference to ratios of this metric and define how a person stands in proportion to others. Interactions comprising moderate levels of MP may simply be considered professional `business-like' relationships, in which actors are often concerned with achieving similar ratios as others. However, within strong MP relations, distinctively oriented towards personal achievement and self-interest with a focus on efficiency and efficacy, members engage in rather intensive cost--benefit analyses, and mainly compete to rationally maximize their personal gains, while minimizing their losses. Fairness of resource distribution is determined by the `equity' principle, with the individual payoffs supposed to proportionally reflect one's contributions, and decision making is individually performed while being coordinated in the background through the free-market mechanism. Hence, the meaningful social interactions under this model tie in with ratio scales \citep{Fiske:1992}. One should note that the continuous rational cost--benefit analyses and computations within MP are quite mentally involved, making this model the most sophisticated and operationally intensive form of sociality \citep{Fiske:1992, Giessner:2010}.

As research has shown (e.g., \cite{Fiske:1992}), an important observation involves the increasing level of complexity among these four sociality forms, according to the ordering:
\begin{equation}\label{formhierar}
\text{CS} \to \text{AR} \to \text{EM} \to \text{MP} \ .
\end{equation}
In other words, the degree of sophistication of meaningful social interactions increases as one moves from the `simplest' form of sociality, CS, towards the `most complex' form, MP; each model encompassing all or most of the relations and operations defined in the preceding models, while including new, previously undefined ones.\footnote{It is important to emphasize that the four basic sociality forms describe socially \textit{meaningful} interactions, which are accepted and adhered to by the involved parties. Using or abusing one another within a relationship is characterized by the \textit{asocial} relation, which does not correspond to a socially meaningful interaction, and is, thus, outside the scope of current treatment.} This observation is argued to correlate with the enhancing cognitive capacity in the course of evolutionary history of humans and other species.\footnote{Similar ordering in externalization of the four relational models is also observed in children growing up, reflecting their enhancing cognitive development with age (see, e.g., \cite{Fiske:1992}, Tab. 1).} Indeed, although forms of CS are observed within simpler lifeforms (e.g., aggregations of unicellular organisms), the more advanced AR sociality form is known to additionally occur within social insects (e.g., bees and ants). EM is observed within the more evolved species (e.g., social birds and mammals) in addition to CS and AR,\footnote{In fact, AR dominance hierarchies are clearly exhibited by vertebrates.} whereas the most sophisticated form MP --- being cognitively demanding --- appears to be exclusive to humans.

As previously mentioned, to address the complexity of social interactions, it is necessary to combine these four primary sociality schemata within any meaningful social context. The emerging variety in social relations is then determined by two distinct factors: 
\begin{itemize}
  \item \emph{Intrinsic factors}: personal preferences of actors in employing varying levels of each model in different domains of their interactions, based on their individual traits, personality, and character;
  \item \emph{Extrinsic factors}: cultural attributes, norms, and values, prescribing implementation of the appropriate kind and intensity of sociality forms within a specific group and social interaction context.
\end{itemize}

Exploring the combinations of relational models is the subject of metarelational models framework \citep{Fiske:2012}, which considers implicitly or explicitly represented configurations forming structures through which people coordinate their relationships. Such configurations are emotionally imbued, morally evaluated, and can be motivationally directive. Considering relational models as `syntactical' for conjoining relationships, metarelational models are specifically concerned with the combinatorics of sociality forms that inform, entail and~/ or preclude one another (see also \cite{Bolender:2015}). Concerning the number of involved actors and social relationships, as well as kinds of relationships, the original framework describes a heuristic taxonomy to categorize various possible metarelational models. Specifically, a configuration of two sociality forms between two actors is categorized as ``Type~1'' (the `simplest' combination), whereas various possible relational configurations among three actors are considered as ``Type~2'' through ``Type~6'' \citep{Fiske:2012}. As further elaborated in Section~\ref{Funrel}, in the current treatment, we are chiefly interested in a \textit{generalized} version of Type~1 metarelational model, by including all four sociality schemata within the dyadic interaction.

The (meta)relational models theory and its applications have been influential in a multitude of fields of research. \cite{Mossholder:2011} utilized the theory to examine the linkage between various HR systems, their respective relational climates, and the induced impact on employee helping behavior. The relational aspects of knowledge sharing behavior within organizations were studied by \cite{Boer:2011}, and more recently by \cite{Arendt:2022} within the context of team cooperation, who demonstrated the importance of alignment between the models employed by stakeholders for their willingness~/ effectiveness of, respectively, knowledge sharing and cooperation. Similarly, the necessity of matching the employed relational models between actors within the context of moral leadership was investigated and confirmed by \cite{Giessner:2010} and \cite{Keck:2018}. Further connections between relational models and morality were also studies by \cite{Rai:2011} and \cite{Simpson:2016}. The theory has been applied by \cite{Bridoux:2016} to analyze boosting joint value creation by stakeholders, allowing for a nuanced treatment of the (broadly-defined) ``relational'' and ``transactional'' approaches to stakeholder management. Moreover, a utilization of the theory for the optimization of supply chain management has been explored, e.g., by \cite{Lejeune:2005} and \cite{Blomme:2014}. The importance of the theory in enabling growth of peer-to-peer sharing (relevant to `sharing' economy) has been emphasized in the work by \cite{Stofberg:2021}. 

Furthermore, creating value in an environmentally-friendly and socially-responsible manner constitutes an increasingly urgent topic on the agenda of many (global) firms and governments, and is emerging as the preferred approach for conducting sustainable business. In many countries, Corporate Social Responsibility (CSR), Environmental, Social and Governance (ESG), and corporate citizenship are highly encouraged, and at times even mandated by the law to various degrees. Early (fragmented) research on this topic primarily focused on including the ecological aspect within the economic development;\footnote{See, e.g., \cite{Carter:2008} for a literature overview.} nevertheless, it became apparent that the macro-viewpoint sustainability must simultaneously address three (interrelated) dimensions, namely economic performance, environmental~/ ecological stewardship, and society (e.g., \cite{Sikdar:2003, Goncz:2007}).

The Triple Bottom Line (TBL) paradigm \citep{Elkington:1998, Elkington:2018, Slaper:2011}, encompassing the three pillars of \textit{Profit}, \textit{People}, and \textit{Planet} (3Ps), simultaneously captures the economic, social, and environmental aspects of sustainable value creation. This hybrid framework has been demonstrated to successfully combine and balance the competing social~/ ecological purpose, traditionally linked with non-profit entities, with the economic purpose, traditionally associated with for-profit organizations \citep{Carter:2008, Wilson:2013}. It suggests organizational activities which may enhance firms' longterm economic and competitive advantage along with positive considerations for society and natural environment. 

TBL has been applied in a vast variety of studies.\footnote{See, e.g., \cite{Alhaddi:2015} and \cite{Tseng:2020} for comprehensive reviews.} The supporting facets of TBL regarding organizational risk management, transparency, strategy, and culture, as well as their interrelationships, were examined by \cite{Carter:2008} through a comprehensive literature study. In particular, they concentrated on the areas of overlap among economic, social, and environmental performance, and developed an influential framework demonstrating how the three goals could be balanced to improve sustainability (with an emphasis on supply chain management). The relationship between corporate governance~/ culture, and the TBL sustainability performance, through the lens of agency theory and stakeholder theory was further investigated, e.g., by \cite{Hussain:2018}. The importance of measuring and reporting TBL within the context of CSR~/ ESG has been emphasized by, e.g., \cite{Goel:2010}.


\section{Method}\label{Funrel}

In this section, the formalism for an analytical approach to the described optimization problem is introduced. We begin by developing a mathematical framework for metarelational models theory, in order to meaningfully parametrize and interrelate its qualitative properties. Among other considerations, insights from the original theory are utilized to derive a universal analytical metarelation for the configurations of sociality forms, to constrain the free parameters, as well as to capture model's overall quantitative behavior. Subsequently, we propose the concept of \textit{Social Relations Portfolio} (SRP) management, and present a procedural methodology enabling the identification of corresponding efficient portfolio, thereby solving the optimization problem.

\subsection{Analytical Approach to (Meta)Relational Models Theory}\label{quant4elem}

In its original formulation, the relational models theory asserts that all four sociality schemata represent autonomous, distinct, and mutually exclusive structures. In this sense, all four are, hence, postulated to be elementary, independently operating from one another. Within the metarelational models theory, however, actors utilize combinations of the four relational models to define socially meaningful interactions, which results in these schemata to inform, entail and~/ or preclude one another. It is, therefore, important to investigate the metarelational rules governing such configurations; in particular, the potential existence of \textit{universal} metarelational rules, which would invariably apply regardless of the situational context. Just as the case with relational models, metarelational configurations have (to various degrees) also been observed in non-human social animals, as well as in human infants;\footnote{See the discussion and references provided by \cite{Fiske:2012}.} hence, given the apparent underlying evolutionary foundations and the lack of necessity for language, consciousness, or symbolic thought capability, (at least certain instances of) such universality is to be expected. 

\begin{figure}
\begin{center}
\includegraphics[width=0.99\textwidth]{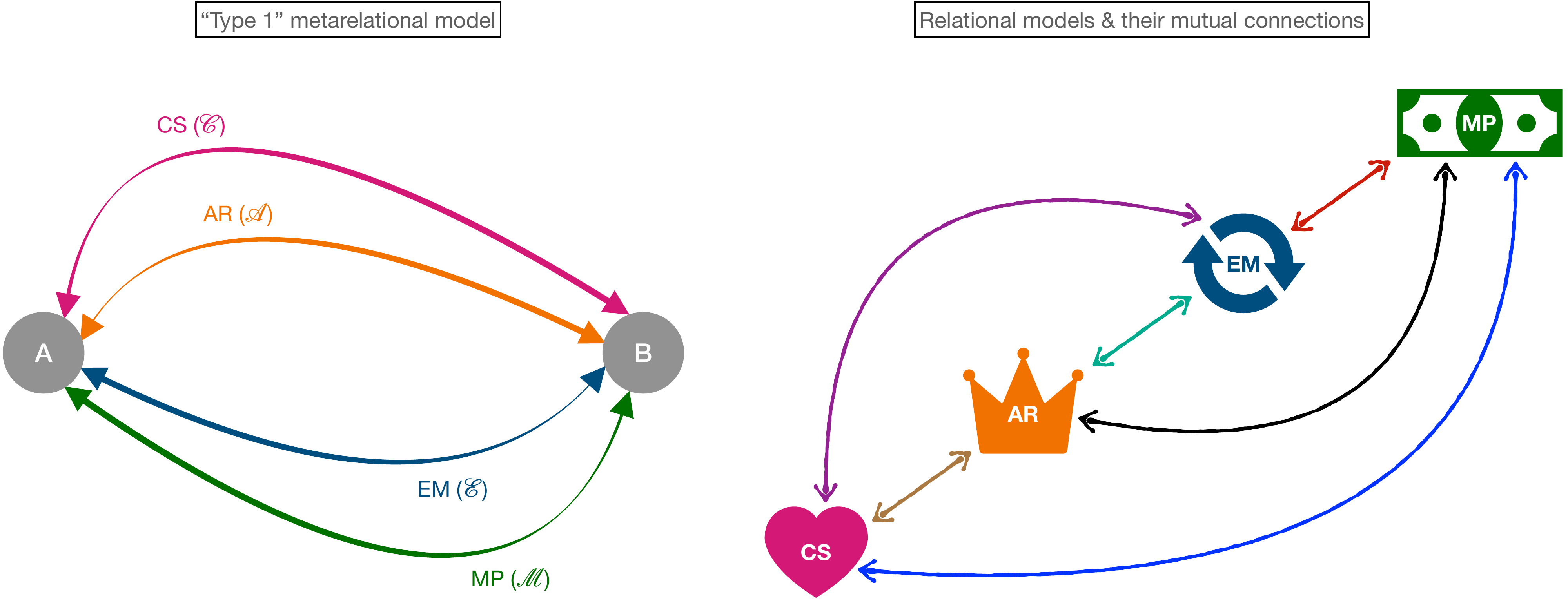}
\caption{\textit{Left}: The generalized form of ``Type~1'' metarelational model \citep{Fiske:2012}. Any dyadic socially meaningful interaction between the actors A and B can be decomposed into a combination of the four sociality schemata CS, AR, EM and MP, respectively parametrized by $\mathcal{C}$, $\mathcal{A}$, $\mathcal{E}$ and $\mathcal{M}$. The (generalized) sum of the latter should add up to 100\% (c.f., \eqref{fundam}), signifying the entirety of the specific interaction. \textit{Right}: The four sociality schemata, with their pairwise metarelational interconnections featured. Taken in pairs, six mutual interconnections between these models are identified, as special cases of their generalized sum \eqref{fundam}.}
\label{Relationsplot}
\end{center}
\end{figure}

A alluded to before, in the current treatment, we are primarily concerned with the combinations of sociality forms between two (generic) actors,\footnote{Here, `actors' are defined in a broad sense; i.e., entities engaging in meaningful social interactions, such as (groups of) people, firms, countries, etc.} which may subsequently be generalized to multiple parties.\footnote{The interesting and important question whether multiple actors' social relations might be the result of a collection of their dyadic relationships or rather one single collective `group relationship' has been contemplated by \cite{Fiske:2012}, but remains unresolved.} By allowing for the possibility to include all four sociality schemata in this combination, we are, thus, considering a \textit{generalized} form of Type~1 metarelational model (left panel of Fig.~\ref{Relationsplot}), discussed in Section~\ref{LitRev}. As we will demonstrate, the generalized Type~1 model contains a single universal (mathematical) metarelation, which fully encodes the interplay among sociality forms within a dyadic relationship.

To begin, let us take a closer look at (mathematical) properties of the aforementioned generalized Type~1 metarelational model (left panel of Fig.~\ref{Relationsplot}). One notes that, within a given dyadic social interaction, it is possible for each of the four sociality models to be employed with varying degrees of strength. In order to meaningfully quantify the strength of corresponding sociality forms, we consider each form to be represented by a dimensionless continuous parameter ranging from 0 to 1,\footnote{For simplicity, in our treatment, we consider the positive and negative forms of each model (e.g., positive or negative reciprocity) to be symmetrical. In future refinements potentially geared towards distinguishing the two, one may allow the parameter domain to vary, e.g., between -1 and 1.}
\begin{equation}\label{pardom}
\mathcal{C} \in [0,1] \ , \qquad  \mathcal{A} \in [0,1] \ , \qquad  \mathcal{E} \in [0,1] \ , \qquad  \mathcal{M} \in [0,1] \ , 
\end{equation}
where $\mathcal{C}$, $\mathcal{A}$, $\mathcal{E}$, and $\mathcal{M}$ respectively symbolize CS, AR, EM, and MP. In \eqref{pardom}, 0 denotes absence of the particular parameter within the relationship; a \textit{null} interaction with respect to the corresponding form of sociality. On another hand, 1 denotes the maximum possible strength with which the sociality form can be exerted in any given situation; for instance, $\mathcal{C} = 1$ may correspond to social communism, $\mathcal{A} = 1$ may represent absolute dictatorship, $\mathcal{E} = 1$ may embody the extreme form of \textit{lex talionis} principle (``an eye for an eye''), and $\mathcal{M} = 1$ may stand for unregulated free-market capitalism.\footnote{In this treatment, we mainly concentrate on the \textit{social} aspects of interactions and their impacts, ignoring any potential political implications.}

The independence and mutual exclusiveness of sociality forms, resulting from their distinct definitions and properties, entails a negative correlation (i.e., `preclusion') among them once they are combined within a certain interaction context (c.f., right panel of Fig.~\ref{Relationsplot}). In a metarelational sense, it is manifestly evident that two or more forms cannot simultaneously be combined with their maximum strengths to define a social relationship, which would otherwise lead to inconsistency and contradictory behavior (e.g., $\mathcal{C} = \mathcal{M} = 1$, corresponding to absolute social communism and capitalism at the same time). In general, a more pronounced sociality form will inevitably suppress the other three within a relationship. For instance, the stronger CS is applied to formulate a specific interaction, the weaker it can be combined with any of the other models. This can be understood, among others, by the diametrical tension between the defining properties of relational models; e.g., equivalency vs.~hierarchy (CS vs.~AR), lack of reciprocity vs.~required balanced reciprocity (CS vs.~EM), and altruism vs.~self-interest (CS vs.~MP). Another example studied on the empirical side considers the tension between a ``transactional'' approach (MP) and a ``relational'' approach (CS, AR, EM) to stakeholder management, and specifically in joint value creation \citep{Bridoux:2016}. It generally reflects the distinction between a rational self-interest motivation and a prosocial motivation \citep{Meglino:2004}; the former exclusively cohering with the individual's personal interest and equity principle \citep{Deutsch:1975}, whereas the latter involving the concern for `the other' and interpersonal factors \citep{Grant:2007}, entailing a negative correlation between (transactional) MP and the other (relational) forms of sociality.

The theoretical and empirical observations described above demonstrate that, within the generalized Type~1 metarelational models, sociality forms universally tend to inform and preclude one another to various degrees, depending on their applied intensities. In other words, combining relational models to formulate a particular social interaction inherently compels internal compromises among them, whereby enhancing one sociality form results in a diminution of the others.\footnote{A practical example is the well-known European Social Market Economy structure, a.k.a. the Rhine Capitalism \citep{Albert:1992, Albert:1993}, which combines free-market economic system (MP) with social safety net considerations (CS), as well as with democratic voting (EM) and governmental leadership and regulations (AR), in a compromising fashion. This is to maximize the overall benefit to the society as a whole, while simultaneously addressing citizens' needs for individuality, belonging to the collective, equal rights, and leadership, among others.} This mechanism appears to be at work in a universal fashion and regardless of the particular situational context. 

From an analytical point of view,\footnote{Throughout this work, we focus on finding the \textit{simplest} mathematical relations and parameterizations compatible with the theoretical and empirical observations, avoiding unnecessary and unilluminating complexities. More refinements of the equations, with additional potential terms and free parameters, may be appropriate for future works if the necessity becomes apparent.} such universal metarelation among the four sociality forms within the generalized Type~1 model may conveniently be parametrized by introducing the (boundary) condition:
\begin{equation}\label{fundam}
\mathcal{C}^\gamma + \mathcal{A}^{\alpha} + \mathcal{E}^{\epsilon} + \mathcal{M}^{\mu} = 1 \ .
\end{equation}
As each of the four sociality parameters can continuously vary between 0 and 1 (c.f., \eqref{pardom}), the universal metarelation \eqref{fundam} enforces the appropriate tradeoffs in their combinations, while explicitly preserving their limit behavior (i.e., only one of the four parameters can simultaneously be equal to 1 within any particular relationship).\footnote{Geometrically, the metarelation \eqref{fundam} amounts to constraining the four sociality parameters to the surface of a 4-dimensional unit hyper-spheroid.} It represents the generalized sum of sociality forms within a generalized Type~1 metarelational model, signifying the entirety of social interaction configuration (i.e., the (generalized) sum of employed sociality forms adds up to 100\%, defining the total nature of relationship).

The power constants $\gamma, \alpha, \epsilon, \mu$ in \eqref{fundam} are positive real numbers, whose exact values are at this point unknown, and are thus introduced as free parameters into the framework. Nonetheless, examining \eqref{formhierar} provides insights into constraints on these power constants. As the degree of cognitive sophistication, and thereby the interaction complexity, increases moving from CS to MP in the order prescribed by \eqref{formhierar}, the sociality forms progressively diverge from CS, with the largest dissimilarity occurring between MP and CS. Utilizing this insight, combined with the fact that CS is the simplest and MP is the most advanced sociality form, one can translate a larger divergence into a greater difference in the respective power constants. Therefore, one arrives at the following inequality for the relational models' power constants:\footnote{It should be emphasized that inequality \eqref{powerconst} excludes the simple linear sum of relational models for configuring a social interaction (i.e., $\gamma = \alpha = \epsilon = \mu = 1$ in \eqref{fundam}), due to the models being inequivalent.}
\begin{equation}\label{powerconst}
0 < \gamma < \alpha < \epsilon < \mu \ .
\end{equation}

\begin{figure}
\begin{center}
\includegraphics[width=.5\textwidth]{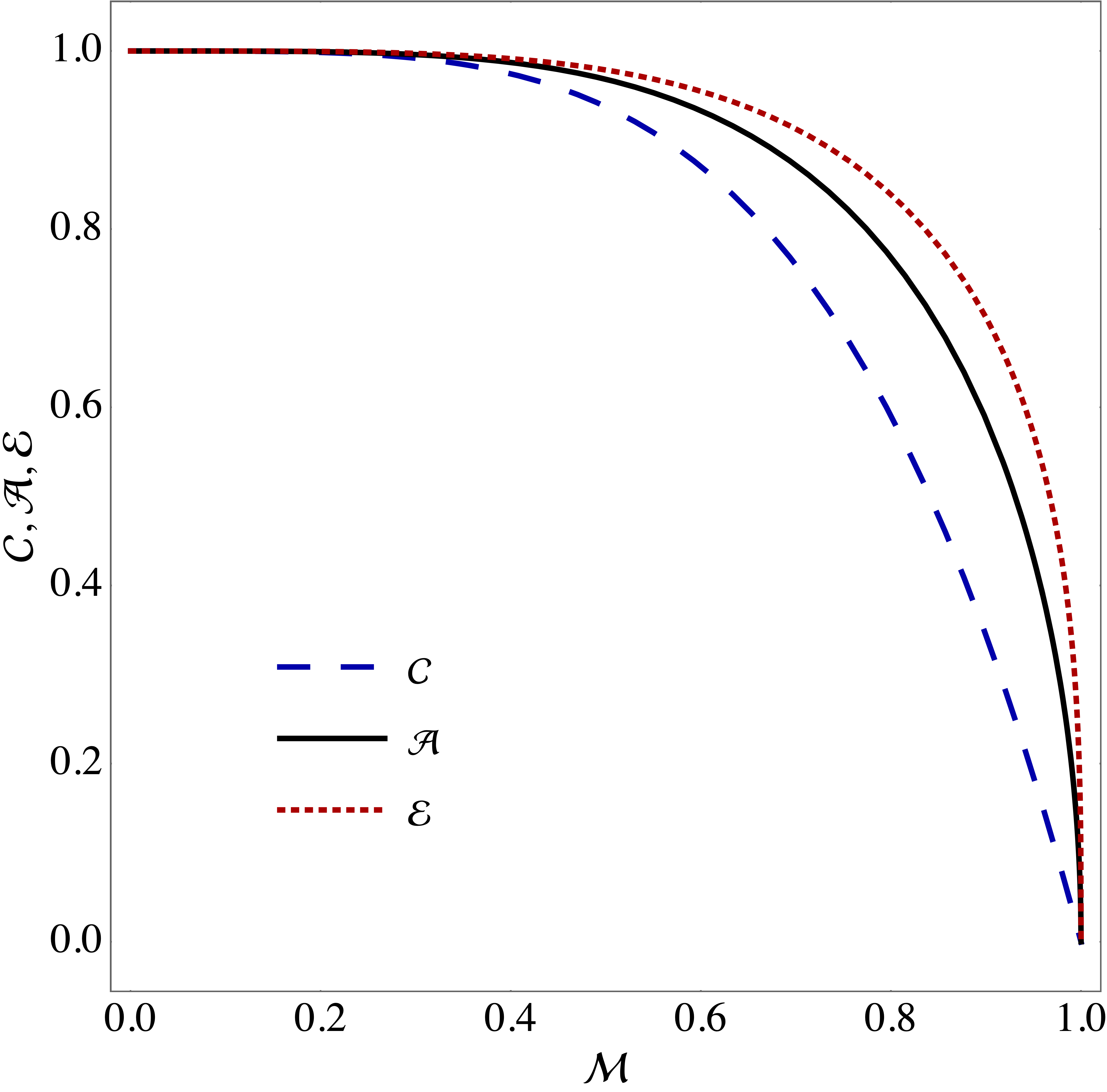}
\caption{The behavior of CS ($\mathcal{C}$), AR ($\mathcal{A}$), and EM ($\mathcal{E}$) as a function of MP ($\mathcal{M}$), given by the universal metarelation \eqref{fundam}, for the illustrative values of power constants in \eqref{PCval}. Note the progressive increase in sensitivity to $\mathcal M$, moving from $\mathcal{E}$ to $\mathcal{C}$, reflecting a larger dissimilarity of these models (i.e., the inherent tension between transactional and relational interactions). In order to display the pure behavior of each parameter as a function of $\mathcal{M}$, the values of the remaining two parameters are set to 0 in \eqref{fundam}.}
\label{CAEM}
\end{center}
\end{figure}

Taking, for illustrative purposes, the pairwise metarelational interconnections between MP and the other three sociality forms as an example (right panel of Fig.~\ref{Relationsplot}), Fig.~\ref{CAEM} displays the behavior of $\mathcal{C}$, $\mathcal{A}$, and $\mathcal{E}$ as a function of $\mathcal{M}$, in accordance with \eqref{fundam}, for the following illustrative values of power constants \eqref{powerconst}:
\begin{equation}\label{PCval}
\gamma = 1 \ , \quad \alpha = 2 \ , \quad \epsilon = 3 \ , \quad \mu = 4 \ .
\end{equation}

In Fig.~\ref{CAEM}, the progressive steepness of the curves, moving from $\mathcal{E}$ to $\mathcal{C}$, demonstrates an increased sensitivity to $\mathcal M$, which, in turn, signifies a larger degree of divergence of their models. Simply put, as the transactional nature of relationship (MP) between two actors enhances, the relational nature of their relationship (CS, AR, EM) accordingly diminishes, with CS being most severely affected followed by AR and EM, as expected. Similar observations can also be deduced for the pairwise metarelational interconnections between all other sociality forms according to \eqref{fundam}.

As mentioned, the power constants currently pose as free parameters within the framework. Controlled experiments need to be carefully designed, in order to quantitatively measure the degree of interdependence of various sociality forms on one another (within dyadic relationships) and gather corresponding statistical data. The \textit{actual} values of power constants can subsequently be calibrated by fitting \eqref{fundam} to the gathered statistical data using regression methods.\footnote{In a similar fashion, the universality of this metarelation can also empirically be verified.}

At this point, it is worth emphasizing that \eqref{fundam} represents the single \textit{elementary and universal} metarelation at the heart of generalized Type~1 configurations. It invariably governs, as an analytical boundary condition, the fundamental interplay among relational models within any meaningful social interaction between two generic actors, defining the total nature of their relationship. Moreover, it leads to the corollary that, within the generalized Type~1 metarelational models, at most only three of the sociality schemata are to be considered as truly distinct and independent; once the intensity of three of the models is chosen (as independent variables) by the actors within their social interaction, strength of the fourth model is then inevitably set (as the dependent variable) by the metarelation \eqref{fundam}.

\subsection{Social Relations Portfolio Management}\label{SRP}

With the (meta)relational models theory on firm mathematical grounds, we turn our attention to introducing the concept of Social Relations Portfolios (SRP) management. In financial economics, it is well-known that for stocks investments combining risk-free bonds with risky securities, one can identify a \textit{unique} optimal combination of securities --- the so-called `efficient portfolio' --- that maximizes returns, given a pre-chosen level of risk \citep{Markowitz:1952, Tobin:1958}. The mathematical optimization approach underlying this discovery, and its subsequent more advanced refinements, have evolved into one of the main tools of asset management at the heart of financial industry.

Taking cues from the financial assets management, in this section we introduce an analogue method for identifying the `optimal' combination of relational models among actors intending to achieve one or more mutually-agreed objectives. In other words, considering the four sociality forms as basic ingredients for constructing a Social Relations Portfolio (analogous to stocks and bonds for an investment portfolio), we provide a procedural methodology for determining the \textit{efficient} SRP, given a mutually-agreed interaction objective (analogous to a pre-selected level of risk). The efficient SRP is, thus, equivalent to the specific configuration of sociality forms which should ideally be adopted by the involved actors who seek to optimize their interaction goal(s). In this sense, the presented methodology addresses the question of how the metarelational configuration within a specific social relationship \textit{ought} to be (from an optimization perspective), as opposed to how it \textit{is} in reality (prescribed by actors' extrinsic cultural rules and~/ or intrinsic personal attributes). Thus, much like financial assets management, this approach enables SRP and stakeholder management, allowing for appropriate corrective actions to be undertaken by the actors in order to steer their relationship toward its optimum, in case (notable) deviations between `\textit{is}' and `\textit{ought}' are detected.

To this end, suppose two generic actors are engaging in a dyadic meaningful social relationship with the purpose of achieving a specific \textit{aligned} goal\footnote{See also the relevant discussions regarding pretense \citep{Fiske:2012}, as well as ambiguity and negotiation of relational models \citep{Lee:2010}.} (e.g., a business deal, mutual cooperation, information sharing, sustainability efforts, etc.) The central question one would be interested to answer is then the very same RQ put forward in Section~\ref{Intr}; i.e.,
\begin{quote}
``\emph{What is the desired combination of the four relational models within a dyadic social interaction, in order to optimally achieve the intended mutual goal?}''
\end{quote}

As mentioned, determining this optimal metarelational configuration is equivalent to constructing the efficient SRP. The latter can mathematically be identified by systematically conducting the following procedural steps:
\begin{enumerate}[a)]
  \item Identify the (set of) aligned objective(s) between actors, facilitated by their dyadic social interaction (see Section~\ref{LitRev} for some examples studied within the literature; e.g., helping behavior, knowledge sharing, joint value creation, supply chain management,~\dots).
  \item If necessary, assign an appropriate measurable proxy to make the identified objective more concrete, and (more directly) connect it to the four sociality forms (see Section~\ref{Applic} for some examples; e.g., productivity as a proxy for profits).
  \item Specify the optimization context for the objective~/ proxy; this typically involves finding a (global) extremum (e.g., \textit{minimization} in case of risk, costs,~\dots, or \textit{maximization} in case of productivity, returns, cooperation,~\dots).
  \item Determine the (graphical) behavior of objective~/ proxy as a function of each sociality form, in the regions of weak, moderate and strong intensity of the latter. This is a crucial step, as it allows for mapping the objective~/ proxy's response to various strengths of the sociality forms. The investigation can be performed, e.g., by thoroughly searching the available literature for relevant experiments and empirical data, or by designing and conducting controlled experiments and gathering the relevant data (e.g., the behavior of productivity as a function of weak, moderate and strong AR relationship).
  \item Cast the obtained (graphical) behaviors in mathematical representations as functions of the four sociality parameters \eqref{pardom}. These functions formally serve as input for the optimization analysis (e.g., a mathematical profile of productivity as a function of $\mathcal A$, for the entire domain $ \mathcal{A} \in [0,1]$).
  \item Add the four mathematical functions together (in a weighted fashion) to obtain the full analytical profile of objective~/ proxy. Use the universal metarelation \eqref{fundam} as a boundary condition (see Section~\ref{quant4elem}) to eliminate one of the sociality parameters as a dependent variable (e.g., the full analytical profile of productivity as a function of only $\mathcal {C,A,E}$, with $\mathcal M$ eliminated using \eqref{fundam} as a dependent variable).
  \item Utilize computational routines to optimize (step~c) the obtained analytical profile (i.e., find the global maximum or minimum, depending on the objective~/ proxy). The corresponding values of sociality parameters (i.e., the coordinates of objective~/ proxy's extremum in the parameter space of sociality forms) determine the efficient SRP, constituting output of the analysis (e.g., maximize the obtained full productivity profile, with the corresponding values of $\mathcal {C,A,E,M}$ defining its efficient SRP).
\end{enumerate}

The methodology presented above enables (mathematical) determination of the desired relative configuration of relational models between two actors, aiming at optimizing their mutually-agreed interaction objectives; it, thus, answers the main RQ proposed in Section~\ref{Intr}. It should be emphasized that this procedure is general in nature, and can be employed for the optimization of any arbitrary concept involving social interaction configurations in a metarelational sense (see also Section~\ref{Implic} for additional discussions). In the following section, as a concrete example, we apply this procedure to the TBL paradigm and (illustratively) identify its efficient SRP, explicitly demonstrating practical and versatile application of the methodology.


\section{Application}\label{Applic}

In this section, as an illustrative application of the developed SRP methodology, we examine the Triple Bottom Line (TBL) paradigm and derive its efficient SRP. As mentioned earlier, TBL encompasses the three pillars of \textit{Profit}, \textit{People}, and \textit{Planet} (3Ps); hence, this comprehensive exploration allows us, among others, to explicitly demonstrate the wide range of applicability of the devised method.\footnote{One should note that the three examined pillars serve as `archetypes' for their respective concepts in a broad sense. For instance, \textit{Profit} generically refers to any interaction objective mainly oriented towards maximizing financial gains, and its derived efficient SRP, therefore, poses as the `typical' optimal sociality mixture for that purpose. Even so, depending on the specific interaction case (e.g., an American-Chinese vs. a German-Qatari profit-oriented relationship) a slightly different efficient SRP might be expected (reflecting distinct personal and cultural characteristics), which is nonetheless anticipated to reside in vicinity of the typical SRP. This is a consequence of the (largely) context-independent behavior of pillars as functions of the four sociality forms (c.f., Figs.~\ref{Prplots},~\ref{Peplots},~and~\ref{Plplots}), as corroborated by the cited literature.} Moreover, it paves the way for a concrete engagement of practitioners in stakeholder and change management with regards to TBL, by ultimately guiding them to measure and monitor organizations' SRP, compare it with the efficient SRP, and perform potentially necessary corrective actions to enhance their CSR~/ ESG efforts.\footnote{We emphasize that the aim of this work is to determine the \textit{optimal relative mixture} of the sociality forms leading to a maximized TBL (i.e., its efficient SRP), and not the \textit{exact absolute values} of the 3Ps pillars themselves. The exact values of maximized 3Ps pillars (e.g., the exact monetary value of maximized profit) heavily depend on the specific context to which the framework is applied (e.g., particular characteristics of the firm in question), and are outside the scope of current treatment.}

For this purpose, we execute the SRP methodology in two stages. First, we consider each of the 3Ps pillars as an independent interaction goal, and apply the procedure to draw conclusions regarding their distinct efficient SRPs. Armed with this knowledge, we subsequently proceed to find the efficient SRP for entire TBL as a single paradigm. In this fashion, illuminating insights can be obtained into pillars' efficient SRPs and their differences, and valuable contrasts can be drawn between focusing on a single pillar and whole TBL. In the process, and specifically for sketching the graphical behavior of pillars as functions of the sociality forms (step~d of the procedure), we mainly rely on the large body of available (interdisciplinary) literature, which are, in turn, predominantly motivated by empirical data and controlled experiments.\footnote{Hence, we do not directly gather~/ use data, or design controlled experiments within this treatment.}

In the forthcoming subsections, it becomes evident that the profiles of 3Ps pillars as functions of the sociality parameters (step~e of the procedure) typically involve competing factors, which become dominant in different regions. From a mathematical point of view, without loss of generality, such behavior may conveniently be parametrized using the (normalized) product of a power and an exponential function:\footnote{A summary of all developed profiles within the text, as well as the employed illustrative values of free parameters, is provided in Appendix for convenience.}
\begin{equation}\label{ParGen}
P (X) = \frac{1}{N} \cbrac{ X^{l} \times \exp \tbrac{-m \pbrac{\frac{X}{1 - k X}}^{n}} } \quad \quad  \pbrac{ X \in \cbrac{\mathcal{C}, \mathcal{A}, \mathcal{E}, \mathcal{M}} } \ ,
\end{equation}
where $P(X)$ denotes any of the 3Ps pillars (\textit{Profit}, \textit{People}, or \textit{Planet}) as a function of the sociality parameter~$X$ (ranging from 0 to 1, c.f. \eqref{pardom}), and $N$ represents the normalization constant:
\begin{equation}\label{ParGenNorm}
N \equiv \int_{0}^{1} X^{l} \exp \tbrac{-m \pbrac{\frac{X}{1 - k X}}^{n}}  dX \ .
\end{equation}
The variables $k, l, m, n$ may freely be adjusted to reproduce the (by the literature motivated) behavior of each of the pillars as a function of any of the sociality parameters; therefore, they pose as the (minimal) set of free parameters necessary for achieving this purpose. Specifically, adjusting $l$ controls the relative strength of power function, which is typically dominant for lower $X$~values. In contrast, the exponential function typically dominates in the higher $X$~regime, with its relative strength being dictated by $m$ and $n$. Finally, $k$ adjusts the border value at the upper limit, $P(X=1)$.

The normalization of \eqref{ParGen} facilitates a meaningful relative comparison among various profiles, as well as the possibility to add them together. Therefore, in analogy with the sociality parameters \eqref{pardom}, the profile functions are also dimensionless, making them general and universally applicable.

\subsection{Profit}\label{Profit}

We begin by following the SRP procedural steps of Section~\ref{SRP} for the first pillar, \textit{Profit} (step~a), in a generic context (e.g., in a firm). To this end, we utilize productivity and performance as the (measurable) profit proxy to which the sociality forms are ultimately related (step~b), and are interested in the maximization of this proxy (step~c). Consequently, the impact of each sociality form on the proxy should be plotted (step~d), and their corresponding mathematical profiles as functions of the sociality parameters constructed (step~e). These latter processes are described below in detail; the identification of \textit{Profit}'s efficient SRP (steps~f~\&~g) is further explicated in Section~\ref{SRP3Ps}.

\subsubsection{Profit as a Function of CS --- ${\mathcal Pr  (\mathcal C)}$}\label{PRCS}

In order to capture the behavior of profit as a function of CS between (a collection of) dyadic actors, first imagine a situation where little CS within the mutual relationship is present ($\mathcal C \sim 0$). In this setting, the altruistic `humane' element is virtually absent from human interactions; group members may not care much about each other's feelings and needs in a collective sense, do not unconditionally help one another, do not share a common identity, and may consequently exhibit a rather `mechanical' behavior. However, as the level of CS within this setting enhances, the potentially cold mechanical interactions start to `thaw', warming up human connections and exerting a positive effect on members' state of mind and motivation, as well as their group sense of belonging. Such increase in friendliness, compassion, and helpfulness can, in turn, translate into an increased collaboration, coordination, and productivity \citep{Mossholder:2011}, and, by extension, larger expected profits \citep{Podsakoff:2009, Chuang:2010}. Simply put, when workers `feel good' about their interactions with their colleagues, they experience a more pleasant and satisfactory workplace, and are (intrinsically) more motivated to perform and deliver.

Nonetheless, a positive correlation between CS and profits cannot endlessly continue. One should bear in mind that a higher level of friendly casual relations within the work environment can also lead to a larger degree of informality, which, beyond some critical point, may start to undermine the professional nature of workplace (in line with the concept of ``too much of a good thing''). It is beyond this critical CS level where professionalism would largely be replaced by informal, forgiving, and careless socialization; therefore, adversely affecting productivity and, by extension, profits. The workplace, then, starts to transform into a rather messy ``country club'', characterized by a large degree of social support but a low degree of performance, where workers `may have fun but not get much done!' Moreover, high CS could result in citizenship pressure and extrinsically coerced altruism (entailing a loss of individual motivation, among others), which are associated with various personal as well as professional costs \citep{Bolino:2013}; these are --- in the collective --- likely to decrease profits. For CS reaching its maximum ($\mathcal C \sim 1$, equivalent to social communism), very little financial profits are expected.

\begin{figure}
\begin{center}
\includegraphics[width=.49\textwidth]{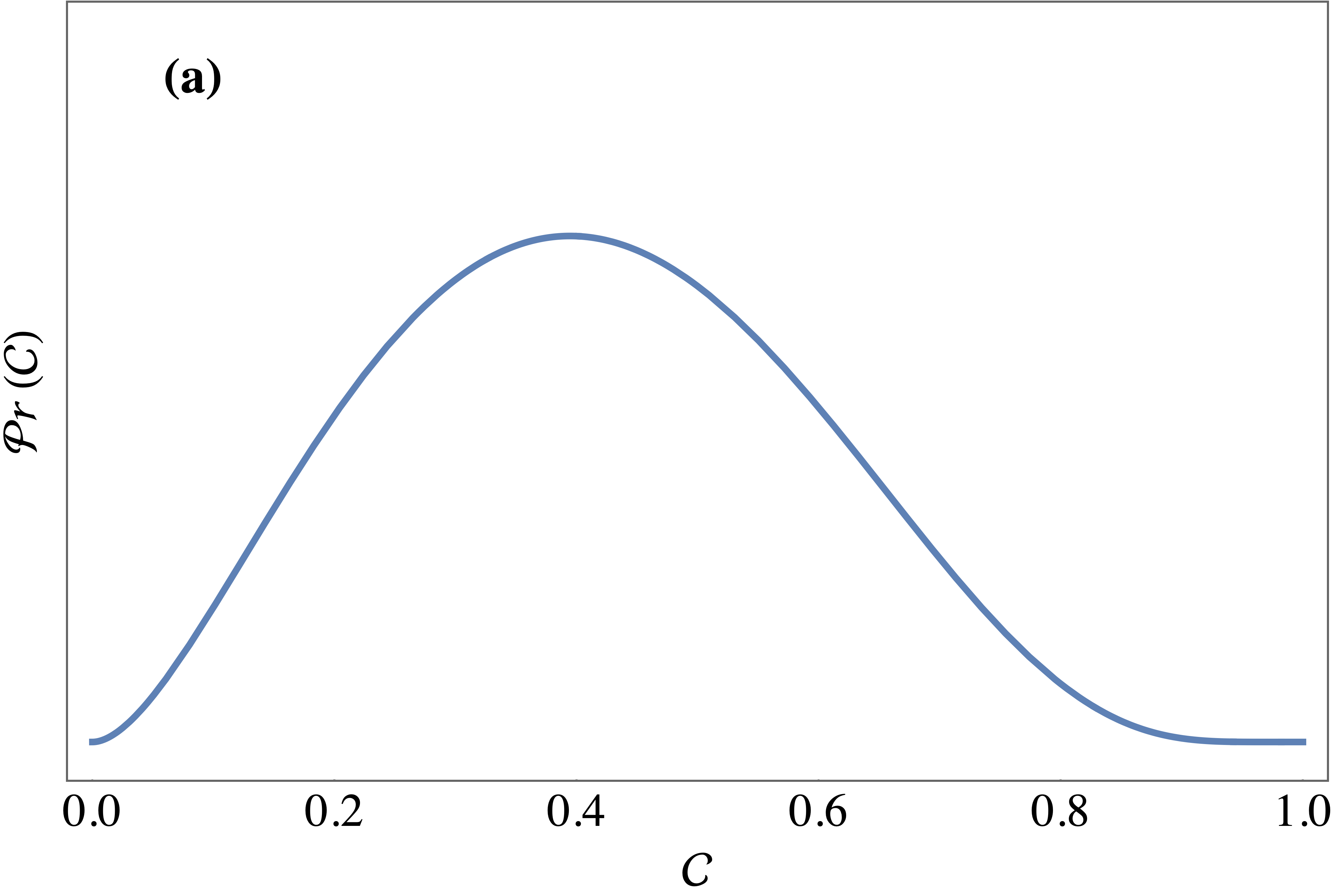}
\includegraphics[width=.49\textwidth]{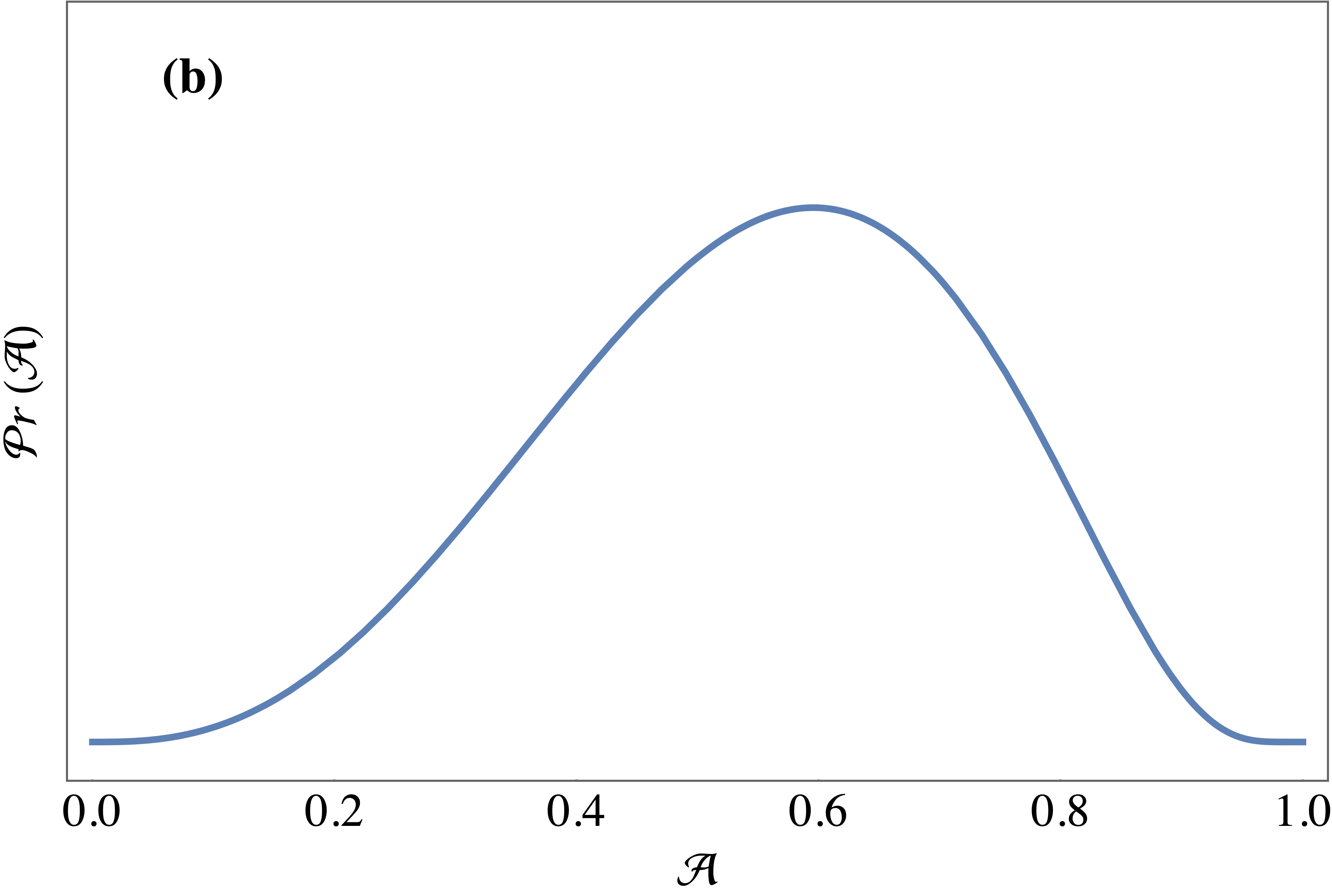}
\includegraphics[width=.49\textwidth]{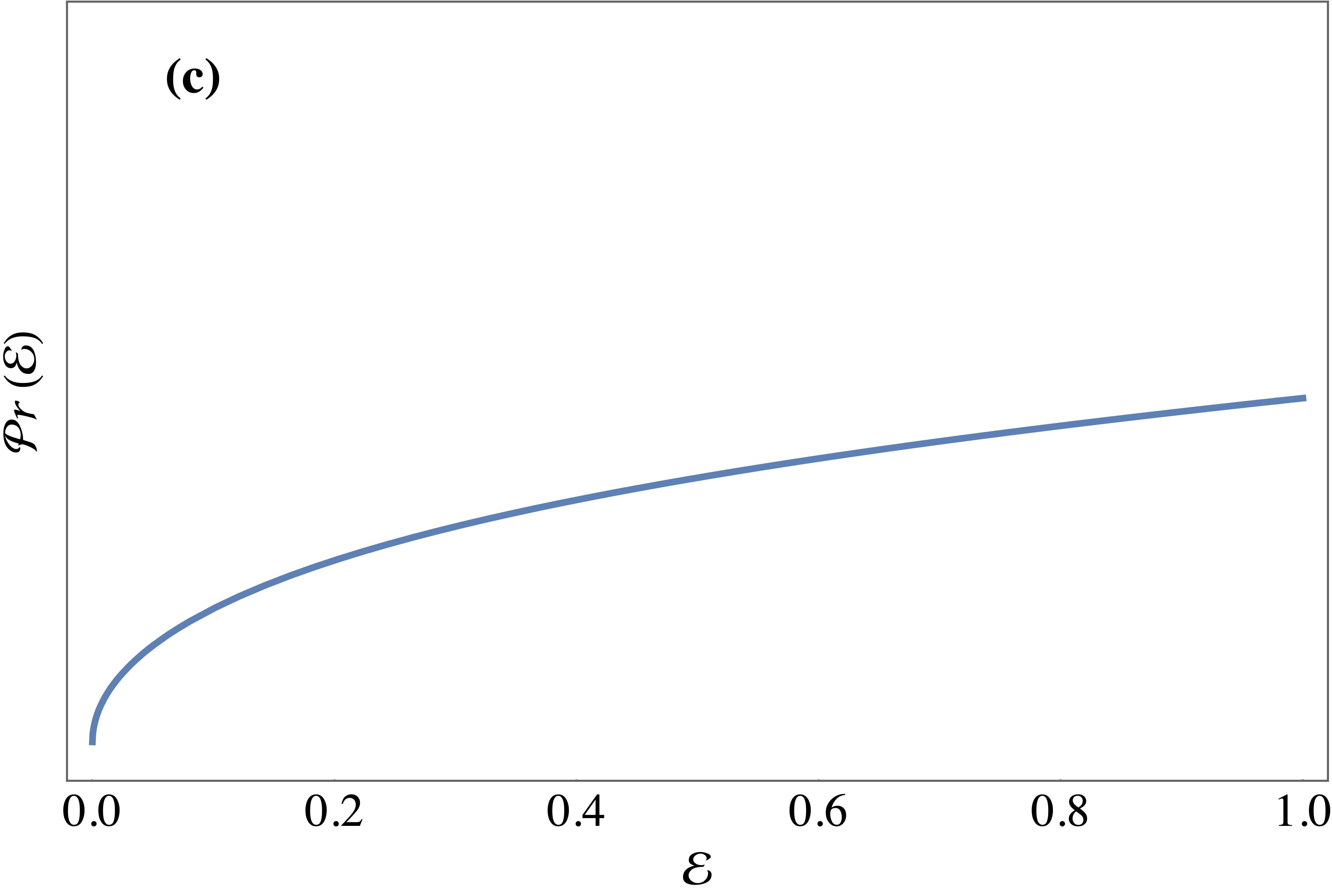}
\includegraphics[width=.49\textwidth]{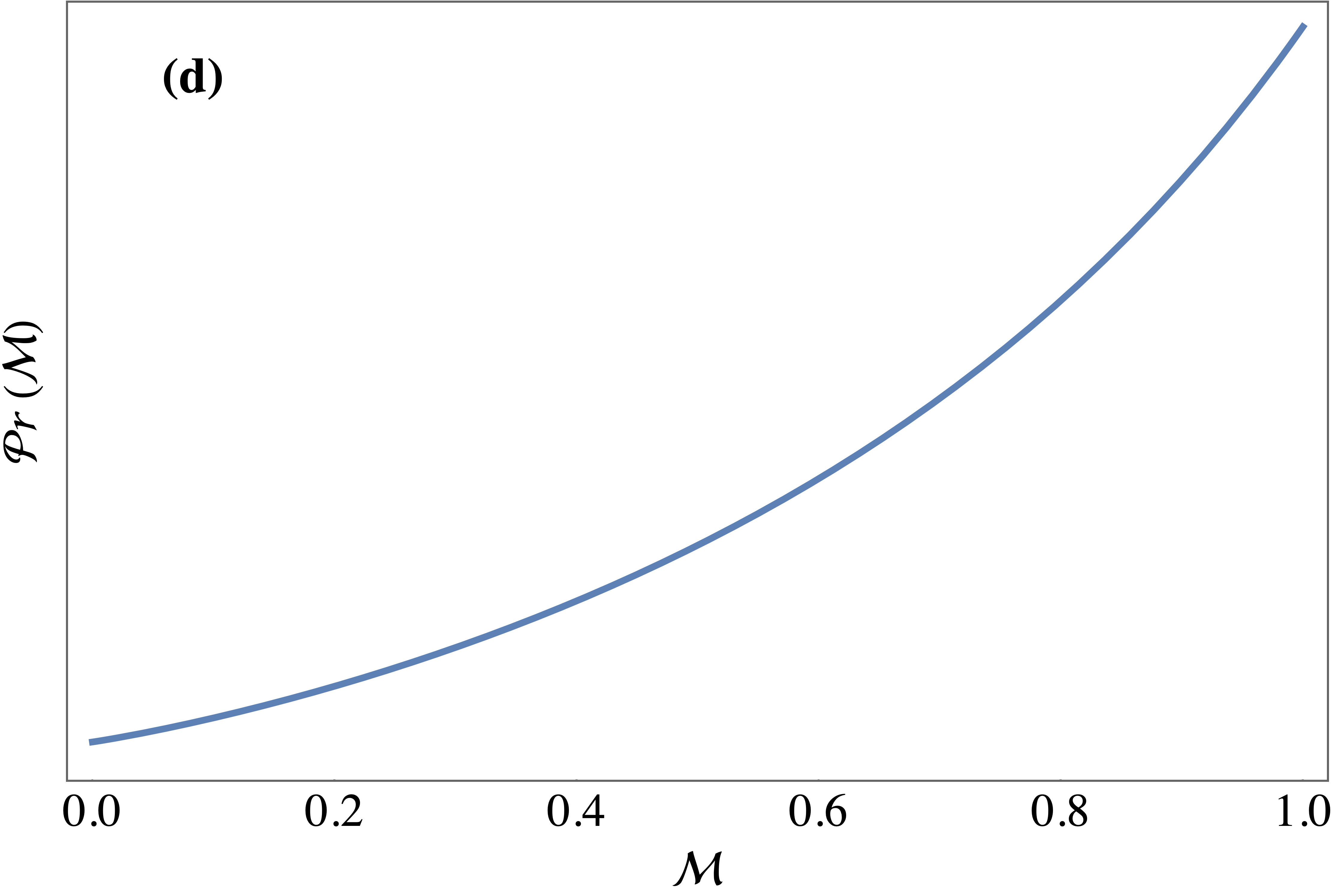}
\caption{\textit{Profit} profile as a function of (a) CS, (b) AR, (c) EM, and (d) MP in arbitrary scale. The location of extrema and inflection points are illustrative. In order to facilitate relative comparison, all curves are normalized.}
\label{Prplots}
\end{center}
\end{figure}

Combining aforementioned observations, the expected behavior of profit as a function of CS is depicted in Fig.~\ref{Prplots}(a). As the level of CS increases within the workplace, an initial (sharp) growth of financial profits is expected (due to a higher degree of social support and more satisfactory work environment), which reaches a maximum at a critical CS level, and subsequently (gradually) declines (due to the ``country club'' mentality and citizenship pressure taking over the workplace); the curve, hence, exhibits a nonlinear behavior \citep{Munyon:2010, Morin:2013}. It is assumed that profits' initial growth is more rapidly realized than its subsequent decline; thereby, setting maximum of the curve within the domain $0 < \mathcal C < 0.5$.

In mathematical terms, the desired profit profile as a function of CS, ${\mathcal Pr (\mathcal C)}$, resulting from the described competing factors, can be constructed using the general parametrization \eqref{ParGen}, by imposing the following constraints on its variables:
\begin{equation}\label{PrCS}
0 < l < m \ , \quad 0 < n < 1 \ , \quad k = 1 \ .
\end{equation}
For illustration, the following values are selected to reproduce the curve in Fig.~\ref{Prplots}(a):
\begin{equation}\label{PrCSval}
l = 2 \ , \qquad  m = 3 \ , \qquad n = \frac{1}{2} \ , \qquad  k = 1 \ .
\end{equation}
As with \eqref{powerconst}, the \textit{actual} values of these variables must ultimately be calibrated by fitting \eqref{ParGen} to (future) empirical data of profit~/ productivity as a function of CS, using regression methods. This observation equally applies to all free parameters of the functions introduced in the remainder of this treatment.

\subsubsection{Profit as a Function of AR --- ${\mathcal Pr  (\mathcal A})$}\label{PRAR}

The (anticipated) profit~/ productivity profile as a function of AR is exhibited in Fig.~\ref{Prplots}(b). This behavior can be understood as follows: with presence of AR little to none ($\mathcal A \sim 0$), there is virtually no leadership within the workplace; workers have no clear instructions or mandates, there may be no (formal) guidelines in place to resort to, chaos may become prevalent, and productivity is expected to be low, whereby profits suffer.\footnote{Even within self-organizing teams some level of leadership and structure is necessary, although more initiating in nature \citep{Stoker:2001}. For an illuminating discussion regarding ``acephalous'' groups and the influence of cross-cutting ties, as well as supernatural perceptions, on regulating human relationships, see \cite{Fiske:2012} and references therein.} With growing AR levels, however, a degree of (legitimate) leadership, structure, and stability is introduced within the environment, clear rules and mandates come into place, and ambiguity and chaos diminishes. This can provide the group members with clarity, guidance, as well as well-defined objectives \citep{Judge:2004}, enhancing their performance and efficiency, and therefore leading to higher financial profits \citep{DeHoogh:2015}.

Nevertheless, as with CS, a positive correlation between AR and profits cannot indefinitely carry on. With an enhancing level of AR approaching authoritarianism or dictatorship, subordinates may start to question the legitimacy of leader's status; initially positive effects of leadership, structure and order are then offset by negative effects of contempt towards the ever authoritarian leader \citep{DeCremer:2006}. In addition, leader's directives and orders may become more arbitrary in nature without clear motivation, explanation or justification, being based on their (unquestionable) personal wishes at a particular moment; consequently, the level of frustration among subordinates may rise, productivity and morale begin to decline, and acts of disobedience, sabotage or rebellion become more widespread \citep{Holtz:2013}. Such destructive dynamics have potentially detrimental impacts on team performance and financial results, as subordinates would be less motivated to follow leader's guidance and focus on their tasks \citep{DeHoogh:2015, DeCremer:2006, Holtz:2013}. In analogy with CS, a critical AR level, hence, seems to exist, beyond which the initially growing profits (due to the effective leadership and structure) reach a maximum and subsequently start to (sharply) decline (due to the hindering dictatorship effects) \citep{Bodla:2019, Lee:2017}. Assuming the latter decline to occur faster than the former growth, maximum of the curve lies within the domain $0.5 < \mathcal A < 1$.

In order to construct a mathematical representation of the desired profit curve as a function of AR, ${\mathcal Pr (\mathcal A)}$, the general parametrization \eqref{ParGen} may again be utilized, albeit with the proper choice of its variables:
\begin{equation}\label{PrAR}
0 < m < l \ , \quad 0 < n < 1  \ , \quad k = 1 \ .
\end{equation}
The following illustrative values reproduce the curve in Fig.~\ref{Prplots}(b):
\begin{equation}\label{PrARval}
l = 3 \ , \qquad  m = 2 \ , \qquad n = \frac{1}{2} \ , \qquad  k = 1 \ .
\end{equation}

\subsubsection{Profit as a Function of EM --- ${\mathcal Pr  (\mathcal E)}$}\label{PREM}

As explained before, equality matching is intimately related to the notion of social exchange \citep{Blau:1964} and balanced reciprocity \citep{Fiske:1991, Fehr:2002}, as viewed by elements of the group. Little presence of EM within the group ($\mathcal E \sim 0$) indicates a lack of (positive or negative) reciprocity among actors; favors are never returned and misdeed is never meaningfully penalized. This can, in turn, cause profound demoralization and demotivation for members to help and do favors for one another (akin to the `fear of exploitation' notion), or to withhold themselves from wrongdoing towards others. Moreover, members may feel not being perceived as equal partners in their interactions or having an equal say. Such lack of meaningful reciprocity within workplace, on the personal as well as organizational level, undermines performance and productivity, and, hence, profits within the organization \citep{Evans:2005}.

Conversely, increasing the level of EM improves the perception of fairness and equal say within the organization. Group members are able to harvest fruits of their favors done towards one another, feel empowered and having a voice within their group, which provides them with a direct incentive to, among others, actively engage in helping behavior \citep{Mossholder:2011}, enhancing their overall productivity. This pattern of increased performance due to improved perception of reciprocity positively influences profits within the organization \citep{Evans:2005}. At a certain level of EM, however, a satisfactory level of justice, fairness, and empowerment within the group is established and accepted by members, beyond which any additional enhancement of EM to its extreme form (e.g., the \textit{lex talionis} principle) does not significantly increase performance or productivity, and thus profits \citep{Cornelissen:2014} --- although it does not necessarily hurts them either.\footnote{The decrease in performance associated with extreme reciprocity, mentioned by \cite{Cornelissen:2014}, is attributed to, e.g., a larger degree of socialization among workers, which in our case is more appropriately captured by CS (c.f., Section~\ref{PRCS}).}

These observations are summarized in Fig.~\ref{Prplots}(c), where the profit curve as a function of EM, ${\mathcal Pr (\mathcal E)}$, is plotted. As explained, with an initially increasing EM within the organization, productivity and financial profits (rapidly) improve, whereas beyond a certain level of EM (assumed within the range $0~<~\mathcal E~<~0.5$), their improvement starts to become more moderate (although not necessarily turning negative), signifying the already established accepted level of reciprocity, fairness, and empowerment within the group.

A mathematical representation of this profile may be constructed using the general parametrization \eqref{ParGen}, with the function's variables appropriately adjusted:
\begin{equation}\label{PrEM}
l < 1 \ , \quad m > 0 \ , \quad n > 0 \ , \quad  k < 0 \ .
\end{equation}
For illustrative purposes, the following values are chosen in Fig.~\ref{Prplots}(c):
\begin{equation}\label{PrEMval}
l = \frac{1}{2} \ , \qquad  m = \frac{1}{2} \ , \qquad n = 1, \qquad k = -1 \ .
\end{equation}

\subsubsection{Profit as a Function of MP --- ${\mathcal Pr  (\mathcal M)}$}\label{PRMP}

Finally, we turn our attention to the behavior of profit as a function of MP. By adopting this sociality form, group elements engage in rather transactional `business-like' relationships with one another. From a personal achievement perspective, they are proportionally rewarded for their individual efforts, while proportionally sanctioned on a lack thereof, echoing the classic homo-economicus behavior and equity principle \citep{Adams:1963}. As such, there is a direct personal incentive to increase one's productivity and performance.\footnote{There is also an indirect motivation to collaborate with others if the cost--benefit calculation of collaboration would exhibit a net personal gain \citep{Ellemers:2004, Haslam:2005}; however, collaboration and coordination would most effectively be covered by a CS relationship \citep{Bridoux:2016, Mossholder:2011}.} By definition, it is evident that little profit is made with low levels of MP ($\mathcal M \sim 0$), whereas, profits start to monotonically grow as MP increases. Assuming agency alignment,\footnote{In order to enforce agency alignment, sophisticated corporate governance standards may be necessary, which is outside the scope of this treatment.} such growth is correlated with the enhanced motivation and productivity of individual members to maximize their gains. In fact, the profit growth is expected to accelerate with each incremental enhancement of MP, where, for the latter reaching its maximum ($\mathcal M \sim 1$, equivalent to unregulated free-market capitalism), the earnings become very large. In particular, one notes that the earned profits at maximum MP (at $\mathcal M \sim 1$), are anticipated to be greater than what could have been obtained by implementing a maximum EM (at $\mathcal E \sim 1$, c.f., Section~\ref{PREM}), reflecting the more sophisticated trade mechanism at work within the MP sociality form (i.e., based on ratios and distribution laws, instead of just simple additions and subtractions as within EM).

The accelerated growth behavior of profit as a function of MP, ${\mathcal Pr (\mathcal M)}$, is demonstrated in Fig.~\ref{Prplots}(d). This profile may mathematically be represented by suitably constraining variables within the general parametrization \eqref{ParGen}:
\begin{equation}\label{PrMP}
l \geq 1 \ , \quad m < 0 \ , \quad n > 0 \ , \quad  k < 1 \ .
\end{equation}
Once more, for illustrative purposes, the following values of variables reproduce the curve in Fig.~\ref{Prplots}(d):
\begin{equation}\label{PrMPval}
l = 1 \ , \qquad  m = -1 \ , \qquad n = \frac{1}{2} \ , \qquad k = \frac{1}{2} \ .
\end{equation}

\subsection{People}\label{People}

In this section, we continue the examination of TBL by concentrating on the application of SRP steps to its second pillar; namely, \textit{People} (step~a). In particular, in order to conceptualize this pillar, we utilize social well-being and stress reduction as its (measurable) proxy within a general setting (step~b),\footnote{Social well-being and stress reduction have extensively been studied within the literature. For reviews on job-related well-being, see, e.g., \cite{DeWitte:2016}, \cite{Makikangas:2016}, and \cite{Nielsen:2017}.} and are interested in maximizing this proxy (step~c). As described below, we need to sketch out the impact of each of the sociality forms on well-being (step~d), and construct the corresponding mathematical profiles as functions of these sociality parameters (step~e). The identification of \textit{People}'s efficient SRP (steps~f~\&~g) is, once more, elucidated in Section~\ref{SRP3Ps}.

\subsubsection{People as a Function of CS --- ${\mathcal Pe  (\mathcal C)}$}\label{PECS}

Communal sharing and altruistic relations have been demonstrated to positively influence people's well-being within their group \citep{Post:2005}. As explained in Section~\ref{PRCS}, an increase in CS is associated with the enhancement of `humane' element within the group, which warms up relations among group members. It results in a more compassionate, friendly, and pleasant community, where people's need to have a sense of belonging is satisfied, they feel happier, are kinder to each other, and (altruistically) care about one another. This has, in turn, a positive effect on members' state of well-being, while reducing stress and anxiety levels \citep{Thompson:2006}.

That being said, as CS approaches its extreme form ($\mathcal C \sim 1$), the initial intrinsically motivated caring character of the group progressively transforms into an extrinsically coerced setting, in which individuals are \textit{required} by the group to act in an altruistic manner, and make compulsory sacrifices (in a general sense) for the sake of the collective at all times. In this regime, actors may perceive a loss of freedom in how to control their relations with others, in line with the aforementioned citizenship pressure, which is associated with various personal costs such as work overload and high job stress \citep{Bolino:2013}. Moreover, the corresponding compulsory extreme altruism, and mandatory self-sacrifice for the sake of the collective, entails a complete forfeit of personal gains of any kind (rewards, benefits of one's efforts, individual incentives), as well as of personal identity, which can significantly demotivate individuals to perform. The (perceived) loss of personal choice, freedom and (materialistic) returns, transformation of ``job resource'' into ``job demand'' \citep{Bakker:2007}, as well as the overall sense of worthlessness and~/ or `invisibility' as an individual completely consumed by the collective \citep{Ahuvia:2002, Suh:2002} can, subsequently, induce very high levels of stress, anger, resentment, depression, and desperation projected towards the group. This is eventually likely to outweigh well-being benefits associated with the earlier CS enhancement, and the resulting toxic dynamics and pressure may then become detrimental to members' state of well-being, potentially to the point of mentally devastating some.\footnote{The harmful effects of compulsory altruism and citizenship pressure on members' well-being can quite vividly be observed, e.g., within cults and cult-like groups.}

\begin{figure}
\begin{center}
\includegraphics[width=.49\textwidth]{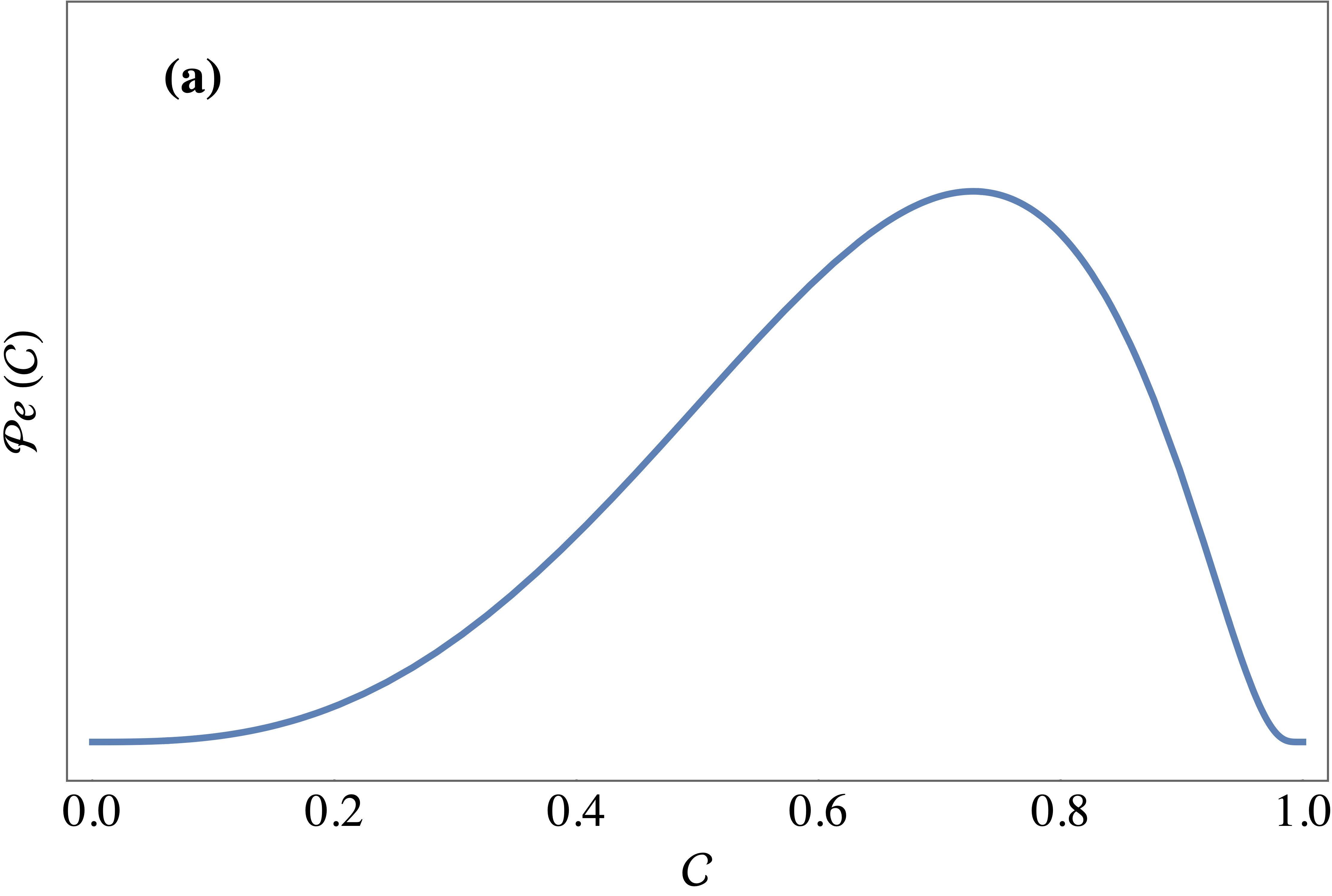}
\includegraphics[width=.49\textwidth]{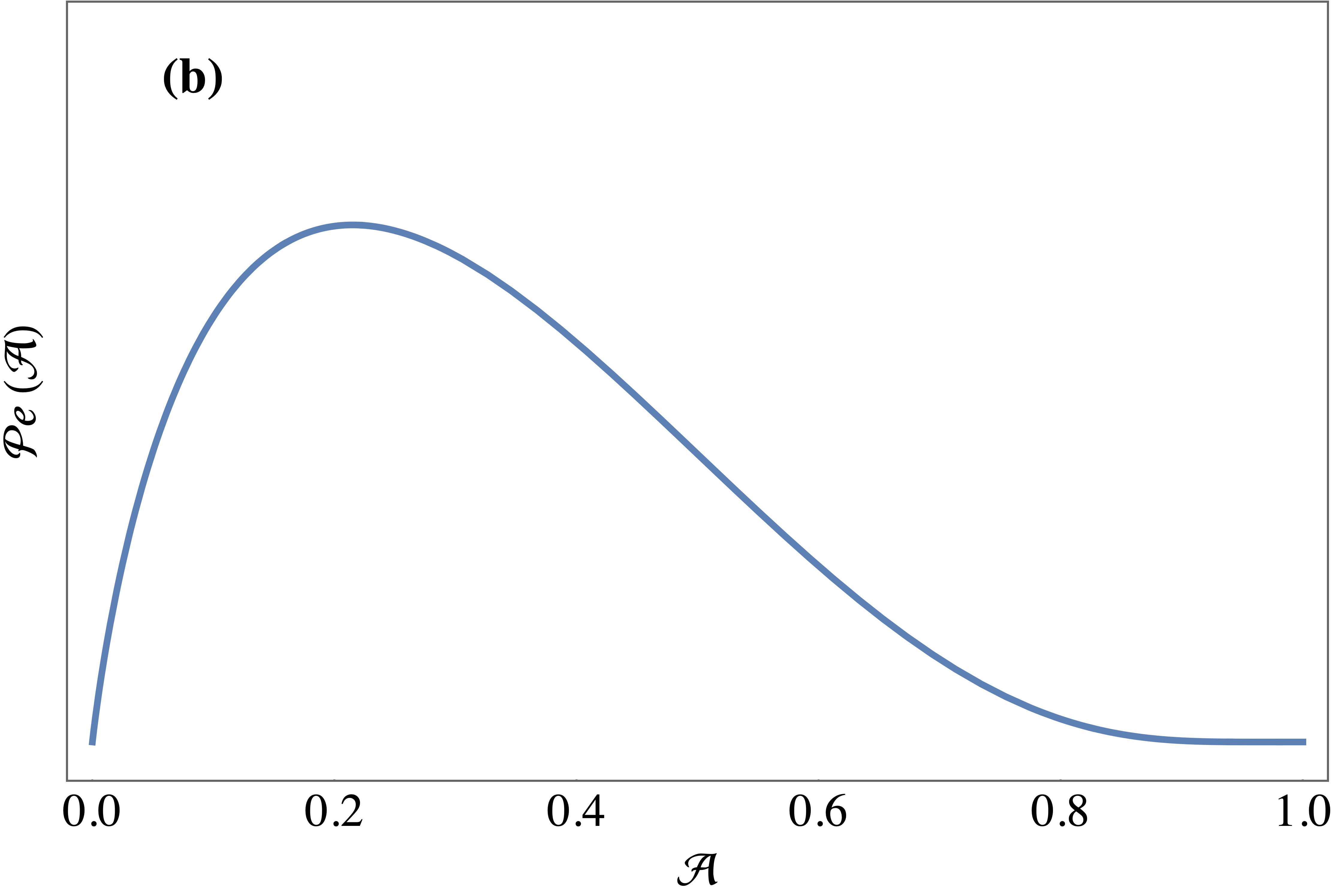}
\includegraphics[width=.49\textwidth]{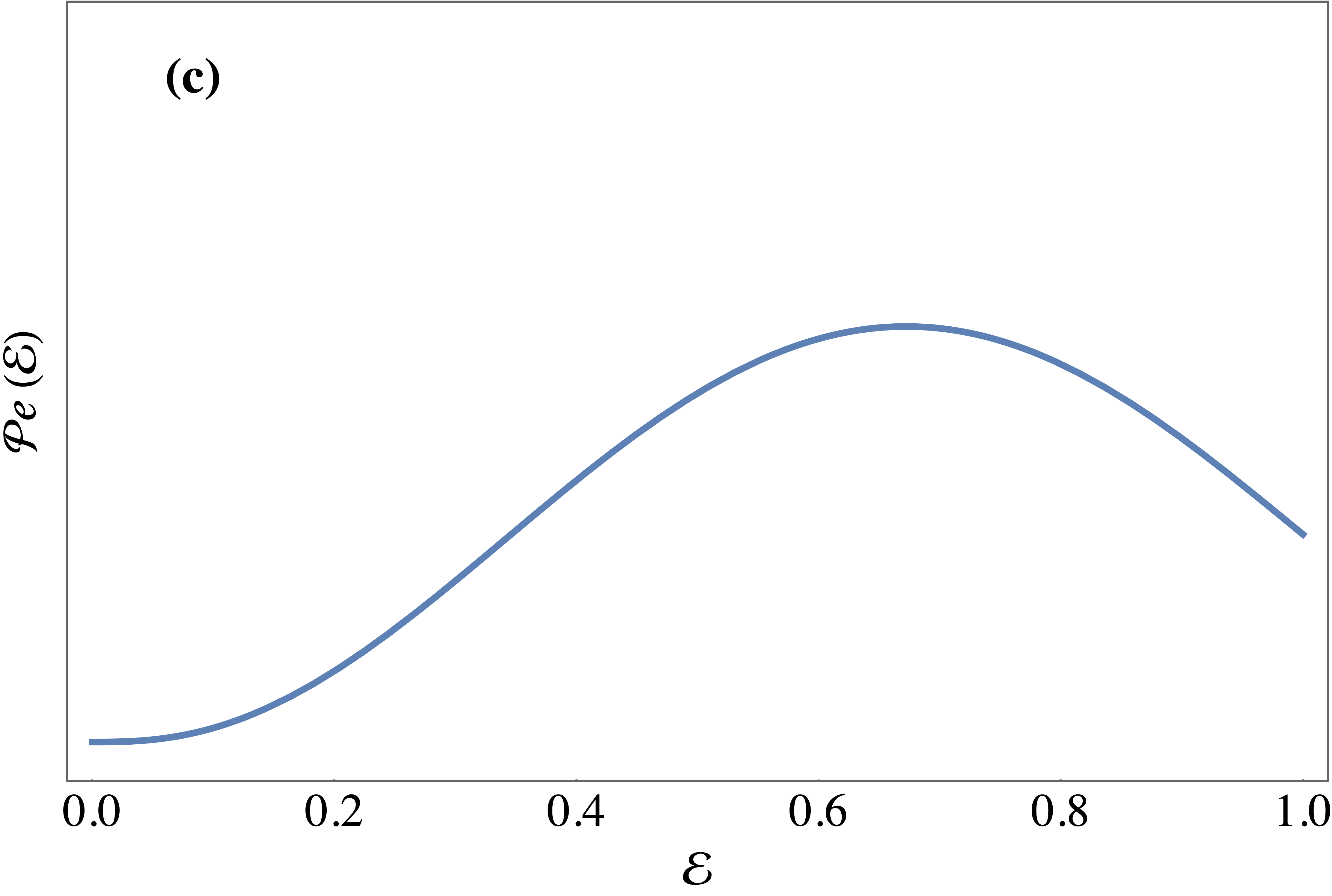}
\includegraphics[width=.49\textwidth]{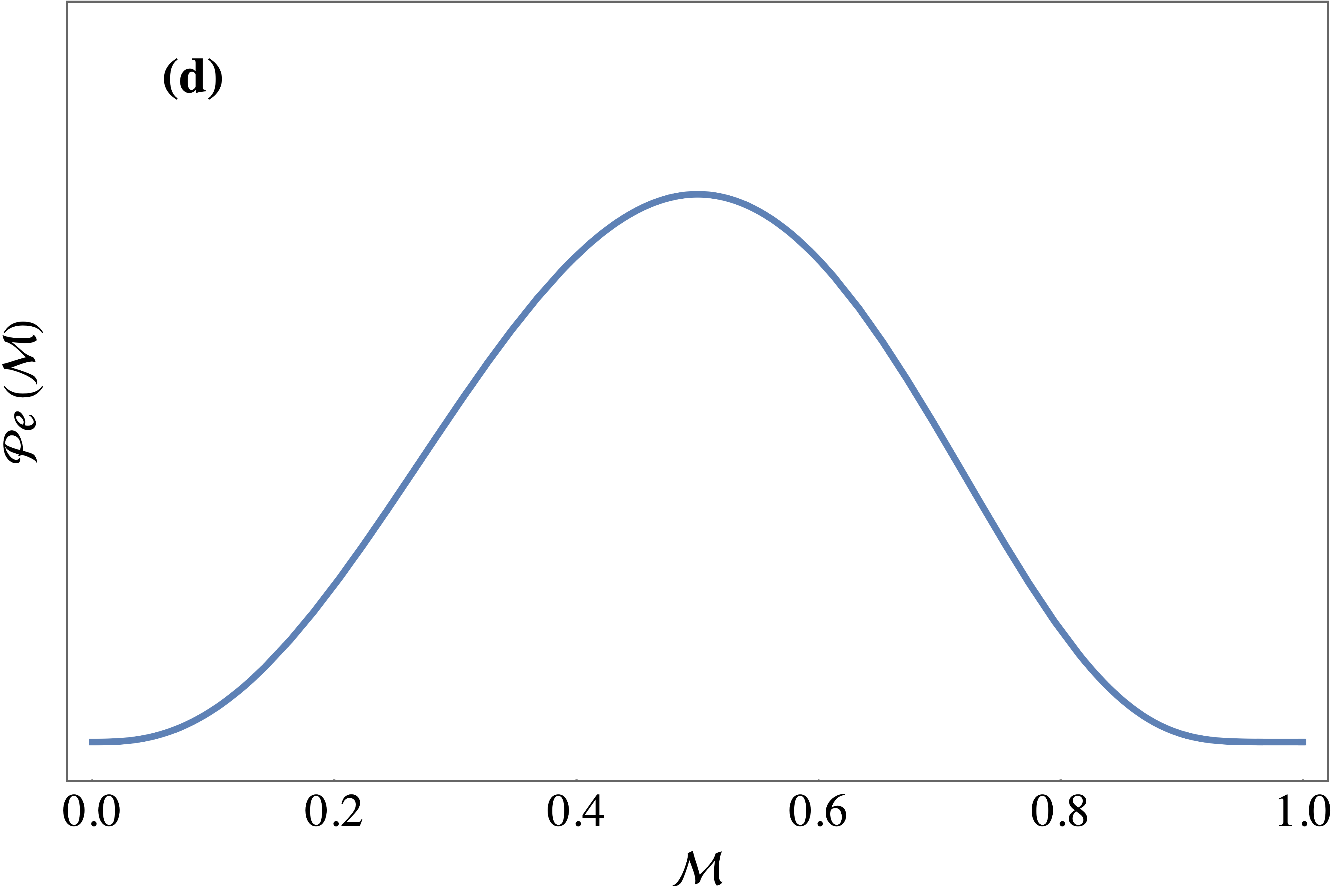}
\caption{\textit{People}'s well-being profile as a function of (a) CS, (b) AR, (c) EM, and (d) MP in arbitrary scale. The location of extrema and inflection points are illustrative. In order to facilitate relative comparison, all curves are normalized.}
\label{Peplots}
\end{center}
\end{figure}

Fig.~\ref{Peplots}(a) exhibits the people's well-being profile as a function of CS, incorporating the abovementioned (competing) factors. A steady enhancement of group members' social well-being is induced by the increasing level of CS (due to the intrinsically induced caring character of the community and the sense of belonging) up to a critical point, after which individuals' social well-being starts to sharply decline (due to the extrinsically imposed loss of choice, freedom, gains, identity, and individuality), once more reflecting the concept of ``too much of a good thing''. Assuming that the latter decline occurs more rapidly than the former enhancement, maximum of the curve lies within the domain $0.5 < \mathcal C < 1$.

The desired people's well-being profile as a function of CS, ${\mathcal Pe (\mathcal C)}$, can mathematically be constructed by utilizing the general parametrization \eqref{ParGen}, with similar constraints on the variables as in \eqref{PrAR}:
\begin{equation}\label{PeCS}
0 < m < l \ , \quad 0 < n < 1  \ , \quad k = 1 \ .
\end{equation}
The curve in Fig.~\ref{Peplots}(a) is reproduced by the following illustrative values:
\begin{equation}\label{PeCSval}
l = 3 \ , \qquad  m = 1 \ , \qquad n = \frac{1}{2} \ , \qquad  k = 1 \ .
\end{equation}

\subsubsection{People as a Function of AR --- ${\mathcal Pe  (\mathcal A)}$}\label{PEAR}

The existence of (a certain degree of) legitimate authority has generally a positive impact on the community as a whole, as it creates order in what might otherwise become a chaotic anarchy. Legitimate leadership, by defining transparent bills of rights and laws, is to a large extent able to eliminate arbitrariness and confusion from group members' interactions, and remove (frightening anticipations of) uncertainty, insecurity, and instability from their community in a streamlined fashion. Predefined rules, guidelines and mandates clarify objectives for individuals, giving them purpose, and the assurance of an undiscriminated and fair treatment free of potential double standards. Moreover, legitimacy of the authority allows subordinates to expect adequate protection and patriarchal care from their leaders, in exchange for recognizing their authority and following their directives. This satisfied need for leadership, certainty, safety, and security under a legitimate leader has been demonstrated to enhance group members' well-being \citep{Seltzer:1988, Lambert:2012, Gaudet:2017}.

On another hand, high levels of AR border with authoritarianism and, in the extreme case ($\mathcal A \sim 1$), with absolute dictatorship. As AR continues to increase, the `protection-for-subordination' social contract between leaders and their followers begins to come under pressure, as subordinates may experience being heavily disadvantaged, and in the extreme case, being subjected to oppression and tyranny. This would cause them to call into question their confidence in the authoritarian leaders and their legitimacy. As a consequence, unrest and stress levels within the group may increase, and the earlier pleasant feelings associated with safety and security under a legitimate caring leadership may give place to loss of autonomy, anger, contempt, and rebellion, which has an overall negative impact on people's well-being \citep{Harris:2006, DeHoogh:2009, Harms:2017}.

The resulting people's well-being profile as a function of AR is displayed in Fig.~\ref{Peplots}(b). With growing levels of AR, there is an initially sharp enhancement of people's well-being (due to order being created in chaos, satisfied need for leadership, safety, and security), which rapidly reaches a maximum at some critical value of AR. As AR passes this critical point and moves towards authoritarianism, people's well-being is reduced (due to loss of confidence in legitimacy of the leader, and the notion of tyranny) \citep{Ames:2007}. Under the assumption that the initial sharp enhancement occurs faster than the later decline, maximum of the curve is expected to reside within the domain $0 < \mathcal A < 0.5$.

One can mathematically represent the people's well-being profile as a function of AR, ${\mathcal Pe (\mathcal A)}$, using the general parametrization \eqref{ParGen}, with a similar choice of variables as in \eqref{PrCS}:
\begin{equation}\label{PeAR}
0 < l < m \ , \quad 0 < n < 1  \ , \quad k = 1 \ .
\end{equation}
The following illustrative values reproduce the curve depicted in Fig.~\ref{Peplots}(b):
\begin{equation}\label{PeARval}
l = 1 \ , \qquad  m = 3 \ , \qquad n = \frac{1}{2} \ , \qquad  k = 1 \ .
\end{equation}

\subsubsection{People as a Function of EM --- ${\mathcal Pe  (\mathcal E)}$}\label{PEEM}

Generally speaking, a positive correlation between the (initially increasing) level of equality matching within a group and its members' well-being has been demonstrated within the social exchange literature \citep{Siegrist:2015, Tsutsumi:2004}. As EM grows, social exchange and the sense of dependable reciprocity strengthens among actors, which favorably influences their perception of equality and fairness. As such, members view one another as trustworthy equal partners who would return favors they owe, building up strong reputations. Conversely, a perceived unjust act by one party is proportionately punished by the affected party. As the group as a whole approves of such retributive measures, the overall sense of accountability and prevailing justice system is fortified within the community. Hence, with a growing EM, notions of dependable balanced reciprocity, accountability, as well as having an equal voice induce a positive effect on members' level of social well-being.

As EM approaches its extreme side ($\mathcal E \sim 1$), akin to the \textit{lex talionis} principle, the degree of anxiety within the community may, nonetheless, start to rise. People then feel obligated to return in kind any and every favor they owe, no matter how small, which might, in turn, make them hesitant to accept favors or help in the first place, fearing to become `indebted' in their relationships \citep{Vecina:2013}. In addition, the fear of unforgiving retaliation towards any misstep, no matter how unintended \citep{Nowak:2006}, might force the actors to constantly look over their shoulders in wary. In such an extreme setting, helping out one another would purely be perceived as a `future investment' by the helper, with an equal return demanded once called upon, echoing reciprocal altruism \citep{Trivers:1971}. Furthermore, the fierce retaliatory nature of \textit{quid pro quo}, `tit-for-tat' and ``an eye for an eye'' doctrines may entail the implementation of potentially inhumane or violent punishments towards offenders (including corporal and~/ or capital punishments), being exactly equal (not just proportionate) both in kind and degree to their committed offenses. Such extreme inimical reciprocity may increase or normalize the overall level of violence within the community,\footnote{Societies implementing capital punishment and~/ or other cruel forms of justice (see, e.g., \cite{Amnesty:2018}) consistently score lower in the societal nonviolence (see, e.g., \cite{GPI:2019}) as well as in the happiness (see, e.g., \cite{WHR:2019}) global indices, as compared with societies not implementing them.} inducing additional (prolonged) societal stress. As a consequence, people's well-being may (somewhat) diminish as EM reaches its extreme limit, but does not necessarily fully vanish.

The aforementioned considerations give rise to the people's well-being profile as a function of EM displayed in Fig.~\ref{Peplots}(c). As EM enhances, people's well-being level also increases (due to the sense of dependable reciprocity and fair accountability). Nevertheless, around some critical point of EM, people's well-being reaches its maximum, and starts to diminish thereafter (due to the induced anxiety associated with reciprocity pressure and potentially higher level of societal violence attributable to harsh punishments), but does not completely disappear. Hence, a nonlinear profile is expected \citep{Moliner:2013}. The turnaround of people's well-being profile is, nonetheless, assumed to occur at large levels of EM; in particular, within the domain $0.5 < \mathcal E < 1$.

A mathematical representation of people's well-being profile as a function of EM, ${\mathcal Pe (\mathcal E)}$, is given by imposing the following constraints on the general parametrization \eqref{ParGen}:
\begin{equation}\label{PeEM}
0 < m \leq l \ , \quad 0 < n < 1  \ , \quad  0 < k < 1 \ .
\end{equation}
Using the following illustrative values, one can reproduce the profile curve displayed in Fig.~\ref{Peplots}(c):
\begin{equation}\label{PeEMval}
l = 3 \ , \qquad  m = 3 \ , \qquad n = \frac{1}{2} \ , \qquad  k = \frac{2}{3} \ .
\end{equation}

\subsubsection{People as a Function of MP --- ${\mathcal Pe  (\mathcal M)}$}\label{PEMP}

The transactional market pricing model is intimately related to the notion of personal achievement and equity. A comparison of this model with CS, for instance, vividly demonstrates the tension between individualism and collectivism; i.e., actors' need for pursuing their personal interests versus their need to belong to a community and be able to successfully function as its members. Accordingly, their social well-being is determined, among others, by delicately balancing the level of transactional nature of their interactions, which, e.g., controls the severity of such tension.

As MP rises, the (instinctual) need for one looking after their own interests becomes progressively satisfied. With professional `business-like' interactions, group members notice that, due to the underlying market mechanism, their investments (financial or otherwise) generally pay off, leaving them `richer' than before, and opening up new doors to previously out of reach opportunities.\footnote{One might say: ``Money may not bring happiness, but no money almost certainly brings unhappiness!''} The group as a whole approves and encourages, towards its members, an attitude of pursuing one's (healthy) ambition and success. This sense of personal achievement and pride, proportionate to one's efforts, capabilities, and investments, is tremendously fulfilling to individuals, and substantially improves their level of well-being \citep{Ahuvia:2002, Suh:2002, Franken:1995, Diener:2009}.

Nevertheless, with the ever increasing level of MP, social relationships among members progressively turn transactional and self-centered in nature, and benign pursuit of personal achievement may transform into malignant greed and unhealthy ambition. The previously applauded attitude towards individual success within the community gives place to a disapproval of inconsiderate behavior, excessive competitive culture, and toxic rivalry dynamics \citep{Jones:1995, Jones:2007}, reflecting the notion of `elbowing' one another and trying to `win at all costs'. Moreover, as a consequence of the underlying market mechanism, those actors whose investments (in a broad sense) have met with success increasingly obtain further access to better and richer opportunities; whereas, those with less luck continue to loose more. In such a system, ``the rich gets richer, while the poor gets poorer'', and the social gap vastly inflates. As MP reaches its extreme limit ($\mathcal M \sim 1$), indicative of unregulated free-market capitalism, all social interactions are framed as absolute cost--benefit transactions, with any notion of helping or caring for `the other' completely consumed by rampant pursuance of (materialistic) personal gains and morbid competition. As a result, stress and tension associated with the relentless cost--benefit analysis and excessive competitive~/ rivalry behavior may approach dangerous levels within the group \citep{Beehr:1998, Tetrick:1987}, with detrimental consequences for people's well-being \citep{Brandts:2008}.

Fig.~\ref{Peplots}(d) summarizes these competing effects, where the people's well-being profile is represented as a function of MP. With a growing MP, people's well-being is initially improved within the community (due to their investments' payoffs, and the fulfilling sense of personal identity, success, and achievement), which reaches its maximum at a critical level of MP, while subsiding thereafter (due to the stress associated with overly self-centered attitude, excessive competition, and inflating social gap). The tension between these effects on people's well-being is assumed to be roughly symmetrical, whereby maximum of the well-being curve occurs around $\mathcal M \sim 0.5$.

The profile of people's well-being as a function of MP, ${\mathcal Pe  (\mathcal M)}$, may mathematically be constructed by utilizing the general parametrization \eqref{ParGen} with the suitable variables' constraints:
\begin{equation}\label{PeMP}
l = m > 0 \ , \quad 0 < n < 1 \ , \quad  k = 1 \ .
\end{equation}
For the sake of illustration, following values of these variables are selected to reproduce the curve in Fig.~\ref{Peplots}(d):
\begin{equation}\label{PeMPval}
l = 3 \ , \qquad  m = 3 \ , \qquad n = \frac{1}{2} \ , \qquad k = 1 \ .
\end{equation}

\subsection{Planet}\label{Planet}

Finally, we explore the connection between relational models theory and third TBL pillar, \textit{Planet}, following the SRP steps of Section~\ref{SRP} (step~a). Environmental impact of human (socioeconomic) activity constitutes one of the most consequential contemporary challenges; hence, it is of paramount importance to include this topic in our discussion. For this purpose, we adopt sustainability and orientation towards longterm impact as the (measurable) proxy (step~b), for which a large body of literature is present. To be specific, we are interested in maximizing this proxy (step~c); therefore, we must comprehend how it is influenced by the four sociality forms (step~d), and find the corresponding mathematical profiles as functions of the sociality parameters (step~e), as elaborated below. As before, the identification of \textit{Planet}'s efficient SRP (steps~f~\&~g) is described in Section~\ref{SRP3Ps}, together with the rest of 3Ps.

\subsubsection{Planet as a Function of CS --- ${\mathcal P \ell (\mathcal C)}$}\label{PLCS}

Communal sharing, in general, exerts a positive influence on the environment \citep{Parboteeah:2012, Tata:2015}, as it is inherently oriented towards the longterm and collectivism, and allows for a reduction of waste, among other effects. Due to the strong caring and helping aspects characterizing this relational model, the community as a whole is inclined to utilize available resources only to the extent necessary in order to improve its members' collective condition. Group members contribute what they can, and take what they need. In this sense, group's production and consumption proportionately balances its overall needs, whereby a sustainable harmony between the group and its environment can be created, contributing to minimizing waste production. Moreover, as environmental habitat of the group is of strategic and vital importance to its collective survival, it is likely to be considered high-priority by its members. The resulting citizenship behavior can, therefore, enhance sustainable or environmental performance \citep{Jiang:2017}. This feature is, for instance, prominently observed within groups of social mammals, which generally tend to reach a natural and longterm equilibrium with their habitat.\footnote{This can be contrasted with, e.g., viruses and non-symbiotic parasites, which, as highly individually- and shortterm-oriented agents, tend to `unsustainably' devastate their habitat (i.e., their host), at eventually great costs to themselves.}

\begin{figure}
\begin{center}
\includegraphics[width=.49\textwidth]{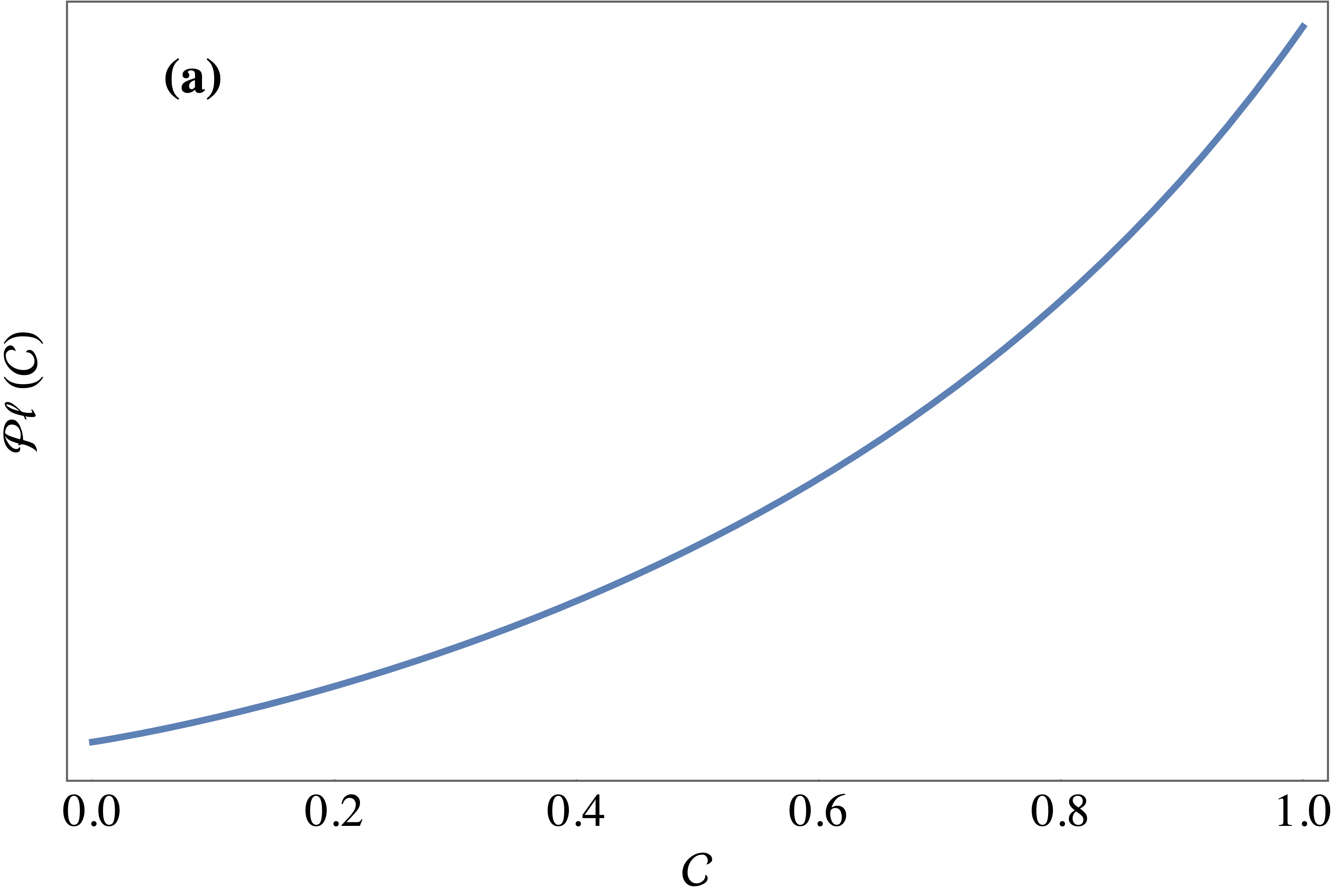}
\includegraphics[width=.49\textwidth]{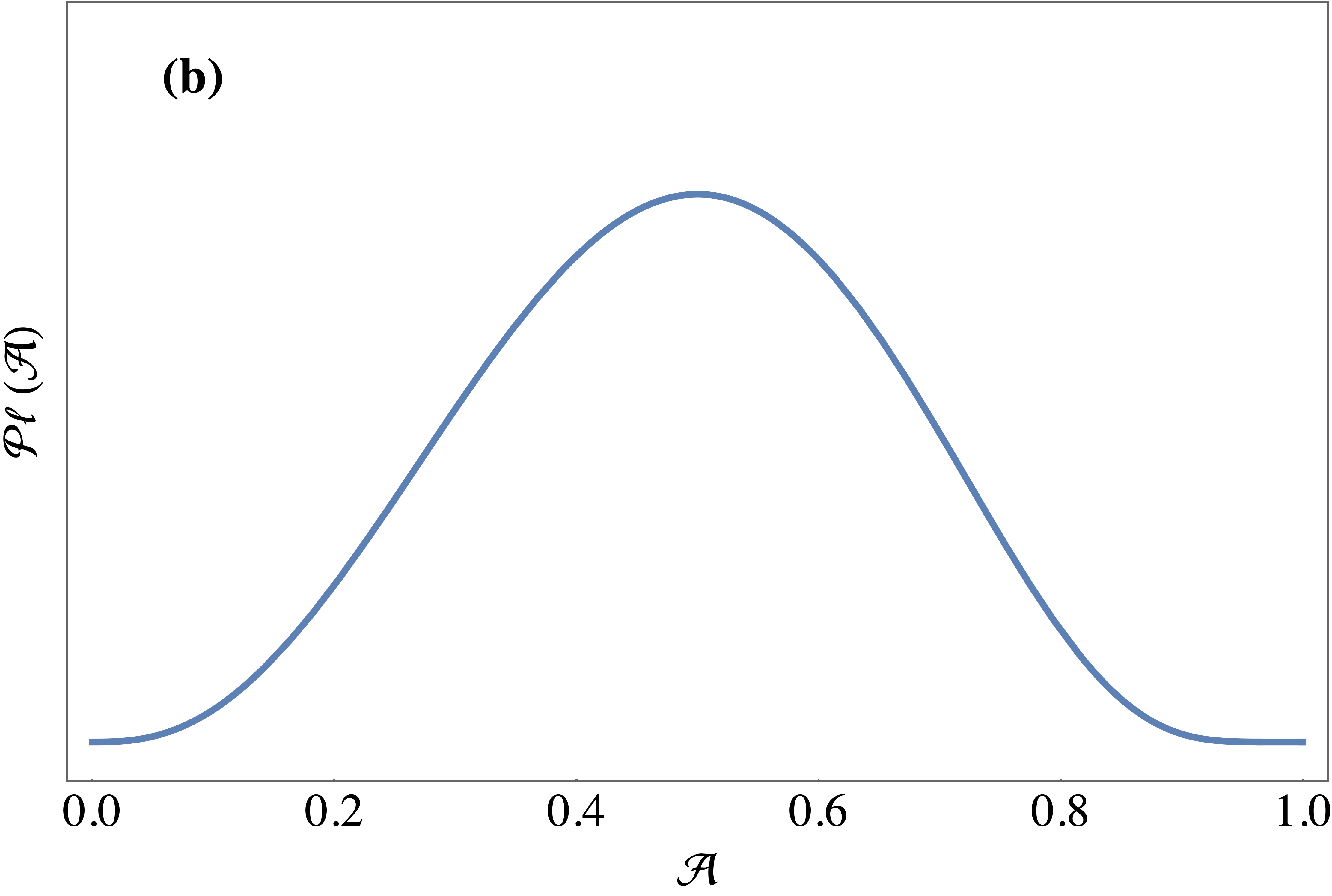}
\includegraphics[width=.49\textwidth]{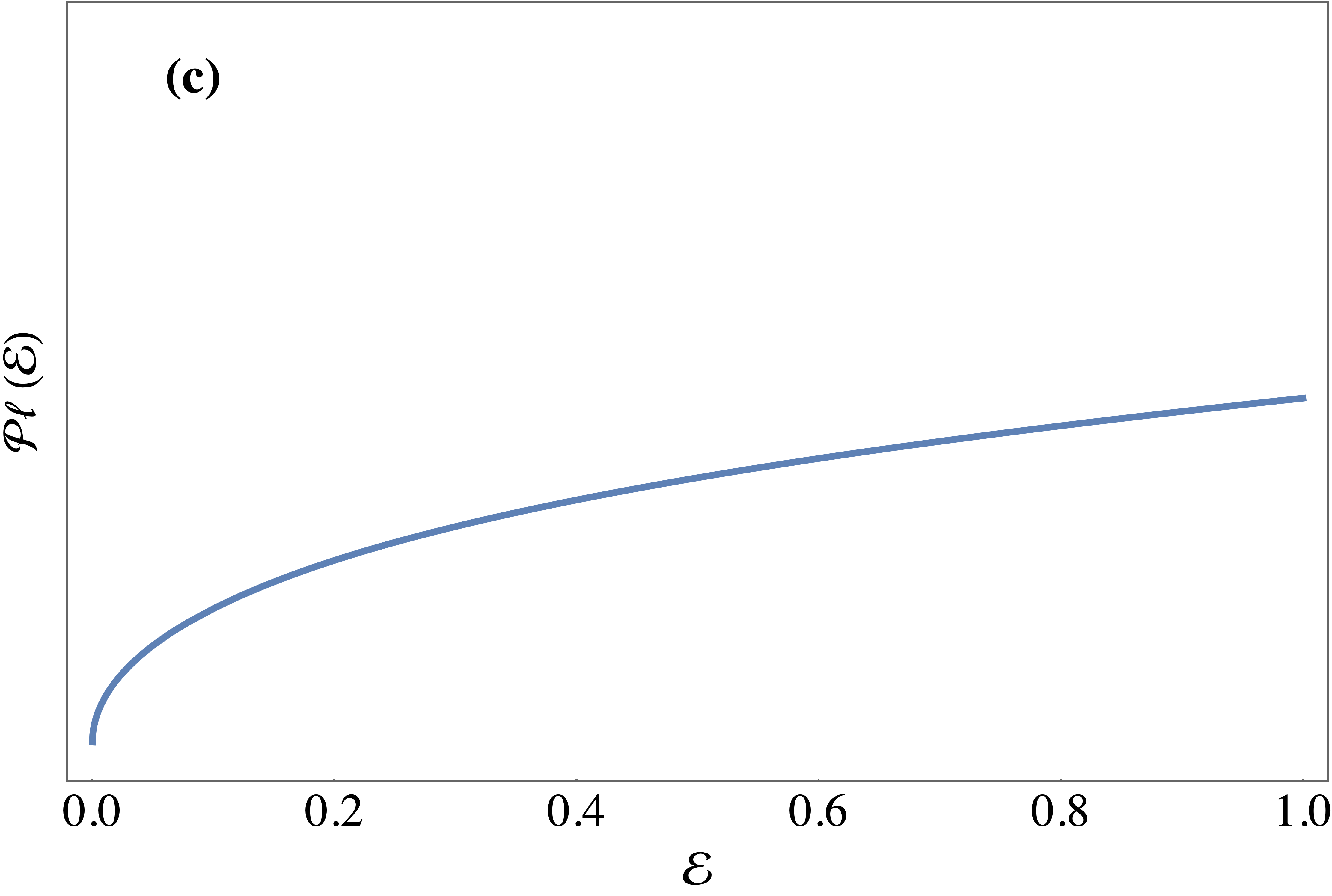}
\includegraphics[width=.49\textwidth]{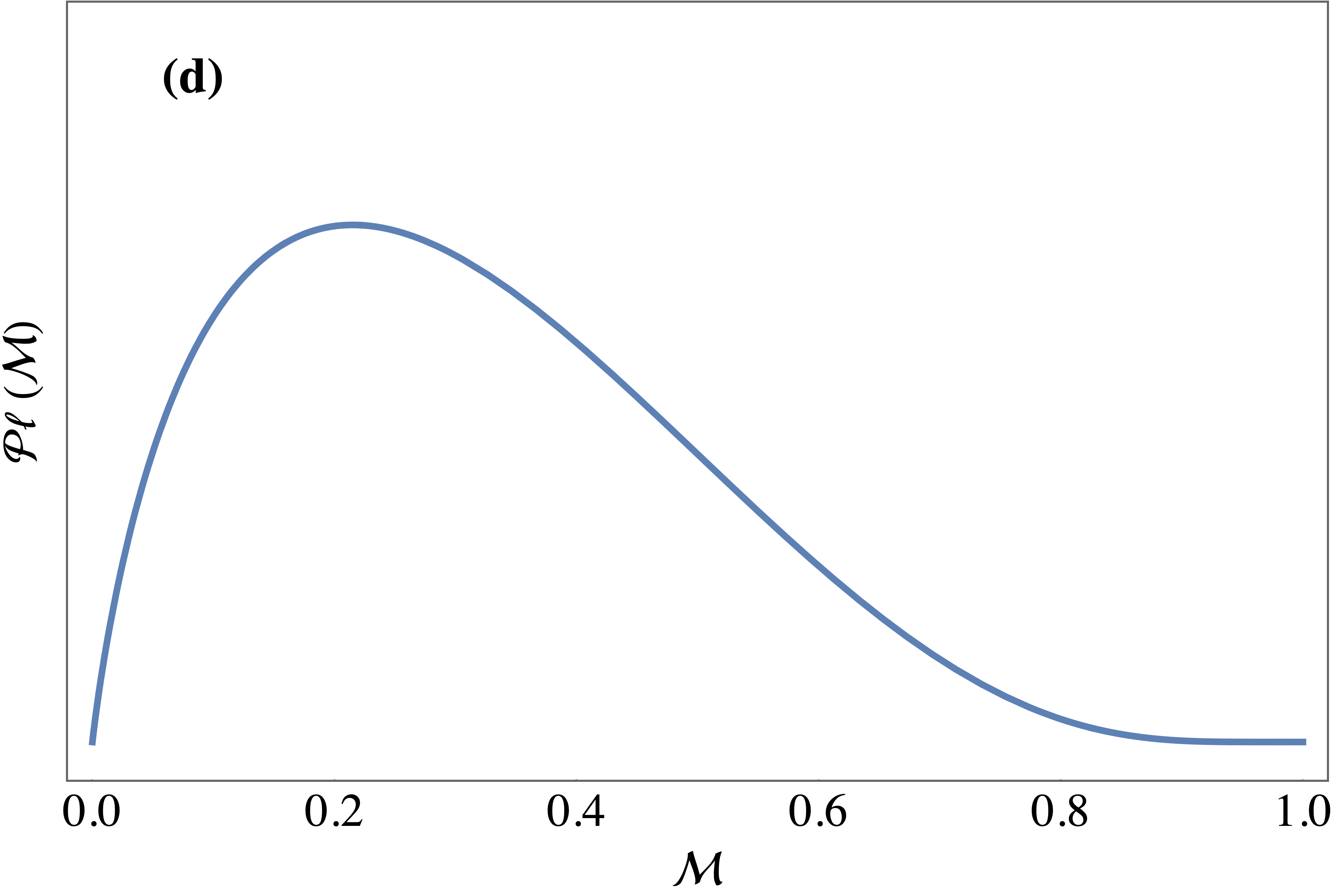}
\caption{\textit{Planet} profile as a function of (a) CS, (b) AR, (c) EM, and (d) MP in arbitrary scale. The location of extrema and inflection points are illustrative. In order to facilitate relative comparison, all curves are normalized.}
\label{Plplots}
\end{center}
\end{figure}

As CS rises, a progressively positive impact on the environment is expected (due to the communal balance of production with group's needs, and tendency for harmonious longterm co-existence with the habitat). This accelerated behavior is demonstrated in Fig.~\ref{Plplots}(a). A mathematical representation of planet's profile as a function of CS, ${\mathcal P \ell (\mathcal C)}$, can be developed by choosing appropriate constraints within the general parametrization \eqref{ParGen}, similar to those in \eqref{PrMP}:
\begin{equation}\label{PlCS}
l \geq 1 \ , \quad m < 0 \ , \quad n > 0 \ , \quad  k < 1 \ .
\end{equation}
For illustration, the curve in Fig.~\ref{Plplots}(a) is reproduced by the following values of these variables:
\begin{equation}\label{PlCSval}
l = 1 \ , \qquad  m = -1 \ , \qquad n = \frac{1}{2} \ , \qquad k = \frac{1}{2} \ .
\end{equation}

\subsubsection{Planet as a Function of AR --- ${\mathcal P \ell (\mathcal A)}$}\label{PLAR}

The impact of authority ranking on environment is anticipated to exhibit a behavior similar to one displayed in Fig.~\ref{Plplots}(b), which could be understood as follows: an initial increase in AR within the group induces leadership, order and structure, and reduces anarchy and chaos. In this fashion, group's (longterm) interests can be identified and prioritized by the legitimate leadership, whereby an effective resource management and allocation may be put in place. Clearly defined instructions and objectives allow members to efficiently perform their tasks and reduce resource misspending. Therefore, an increase in legitimate (inclusive) AR may be associated with an enhanced efficacy, efficiency and structure, and a reduction in futile and wasteful activities, which has a positive influence on the environment.\footnote{See, e.g., \cite{Burns:2015}, \cite{Hargreaves:2004}, and \cite{Quinn:2009}.}

Nevertheless, as AR approaches its extreme level of dictatorship ($\mathcal A \sim 1$), earlier (longterm-oriented) efficiency-enhancing mandates and guidelines may progressively be replaced by arbitrary (shortterm-oriented) desires of the autocratic ruler, potentially diminishing their efficacy or even exacerbating misspending and wasteful activities. The identified priorities and resource allocation would no longer necessarily align with group's longterm and sustainable interests and its environment, but instead primarily reflect those of the dictator. Moreover, as worker autonomy and empowerment are positively correlated with creativity and innovation \citep{Li:2018, Gu:2019}, too much AR might undermine innovative endeavors, potentially suffocating (organic) progress in longterm sustainable developments. Hence, extreme AR and autocracy is expected to have a generally detrimental impact on the environment.\footnote{See also \cite{Bendell:2015} and \cite{Angus:2010}.}

These features are summarized by the curve plotted in Fig.~\ref{Plplots}(b), which demonstrates planet's profile as a function of AR. An enhancing legitimate (inclusive) AR has a positive effect on the environment (due to the effective longterm priority identification and planning, efficient resource allocation and management, and reduction in harmful activities). This positive correlation reaches its maximum around a critical AR level, whereafter the effect of growing AR on the environment turns detrimental (due to priorities and resources diverting towards leader's shortterm desires, and a decrease in efficacy and innovation). Assuming the negative impact of the latter to be roughly symmetrical with the positive impact of the former, curve's maximum is presumably located around $\mathcal A \sim 0.5$.

Planet's profile as a function of AR, ${\mathcal P\ell (\mathcal A)}$, can mathematically be represented by again using the general parametrization \eqref{ParGen}, with the variables' constraints similar to \eqref{PeMP}:
\begin{equation}\label{PlAR}
l = m > 0 \ , \quad 0 < n < 1 \ , \quad  k = 1 \ .
\end{equation}
The curve in Fig.~\ref{Plplots}(b) is given by the following illustrative values:
\begin{equation}\label{PlARval}
l = 3 \ , \qquad  m = 3 \ , \qquad n = \frac{1}{2} \ , \qquad  k = 1 \ .
\end{equation}

\subsubsection{Planet as a Function of EM --- ${\mathcal P \ell (\mathcal E)}$}\label{PLEM}

As previously discussed, equality matching entails reciprocity in a positive sense (returning favors) as well as in a negative sense (retribution acts). Its increasingly prominent presence strengthens mutual trust and honoring of commitments, as well as expectation of accountability, which, in turn, fortify social exchange bonds within the group. Higher levels of EM, thus, lead to an enhancement of members' trustworthiness and reputation as reliable partners, limiting unpredictability in mutual relations, and contributing to the development of strategic alliances. This may then be translated into a positive influence on group's efficiency, as well as diminished risk exposure, reducing the necessity for potentially wasteful mitigatory measures or costly contingency plans. Such alleviation of risk exposure, unpredictability, and fear of exploitation also encourages a longterm-oriented mentality within the group, with members being consciously prone to invest in sustainability of their group and environment. Consequently, a (mild) positive correlation is anticipated between a growing EM and its environmental impact \citep{Konovsky:1994, Paille:2013, Jones:2010}. However, there is no evidence that taking EM to its extreme level within the group ($\mathcal E \sim 1$) might induce a significantly positive (or negative) impact on the environment.

Fig.~\ref{Plplots}(c) displays such a profile, where initially growing EM has a noticeable positive effect on the environment (due to the accompanying efficiency increase, and risk and waste reduction), while becoming more moderate at some critical EM value (assumed within the range $0 < \mathcal E < 0.5$), after which an increase in EM does not lead to a substantial improvement of its environmental impact (although, it does not necessarily induce an adverse effect either).

One can develop a mathematical representation of the planet profile as a function of EM, ${\mathcal P\ell (\mathcal E)}$, by the appropriate choice of variables within the general parametrization \eqref{ParGen}, similar to \eqref{PrEM}:
\begin{equation}\label{PlEM}
l < 1 \ , \quad m > 0 \ , \quad n > 0 \ , \quad  k < 0 \ .
\end{equation}
The following illustrative values give rise to the curve in Fig.~\ref{Plplots}(c):
\begin{equation}\label{PlEMval}
l = \frac{1}{2} \ , \qquad  m = \frac{1}{2} \ , \qquad n = 1, \qquad k = -1 \ .
\end{equation}

\subsubsection{Planet as a Function of MP --- ${\mathcal P \ell (\mathcal M)}$}\label{PLMP}

Environmental impact of the market pricing sociality is sketched in Fig.~\ref{Plplots}(d). Given the sophisticated nature of this relational model, involving ratios, distribution laws, and `price' definition, small levels of MP may be beneficial to the environment, as they introduce measurability of the value of any product or service based on supply and demand (market mechanism), enable rational bookkeeping and accounting for keeping track of transactions, and stimulate commercial sustainability innovations. As a consequence, actors are able to prioritize and adjust their demands according to prices they are willing~/ able to pay, supply can efficiently be streamlined with demand in a measurable and traceable way, and innovative cost-effective solutions may be devised and implemented. In this manner, production and trade can effectively be coordinated, and an efficient market, together with innovative environmentally-friendly endeavors, may deter or outcompete wasteful and polluting conducts. Thus, if properly employed and implemented, small levels of MP can promote sustainable consumption and ecological citizenship \citep{Seyfang:2005}, while aiding `green' manufacturing and production. By aligning economic and ecological purposes for value creation, such environmentally responsible business models may aggressively accelerate innovation in sustainability.

In contrast, an extreme degree of MP ($\mathcal M \sim 1$) results in an unmanageable greed, self-enrichment, and consumerism by the individual elements, as well as the group as a whole. The community descends towards a downward spiral of excessive materialistic consumption, pollution, and waste production; particularly, if the related costs can be offloaded elsewhere or made acceptably low. In other words, the `face value' price tag of products and services may not reflect their `actual' price tag, which would otherwise contain (indirect) environmental costs of their production, as well as longterm effects associated with depleting nonrenewable natural resources \citep{Solow:2014}. As a result, the difference between these two price tags is systemically borne by the planet, with potentially dire consequences for the habitat. In addition, in such a construct, actors' main interests would revolve around shortterm personal wins with immediately measurable (tactical) activities, as opposed to longterm (strategic) and perhaps less tangible sustainability objectives. The sophisticated and continuous cost--benefit calculations, unique to this sociality form, are also costly and vastly energy consuming. Consequently, such an extreme behavior can result in rapid devastation of the habitat and irreversible damages to the environment.

The profile in Fig.~\ref{Plplots}(d) depicts the outlined impact of MP on the planet. An initial enhancement of MP within the community is anticipated to have a sharp positive impact on the environment (due to the effective supply--demand alignment, valuation and bookkeeping, and sustainability innovation), which reaches its maximum around a critical MP level, while turning into a negative correlation afterwards (due to the excessive consumption, pollution, depletion of natural resources, and a focus on shortterm personal goals over longterm environmental interests). Assuming a more rapid initial increase as compared with the later decrease, curve's maximum is expected to reside within the domain $0 < \mathcal M < 0.5$.

A mathematical representation of planet's profile as a function of MP, ${\mathcal P\ell (\mathcal M)}$, is given, once more, using the general parametrization \eqref{ParGen}, with a proper choice of variables' constraints, similar to \eqref{PeAR}:
\begin{equation}\label{PlMP}
0 < l < m \ , \quad 0 < n < 1  \ , \quad k = 1 \ .
\end{equation}
The curve in Fig.~\ref{Plplots}(d) is reproduced by the following illustrative values of these variables:
\begin{equation}\label{PlMPval}
l = 1 \ , \qquad  m = 3 \ , \qquad n = \frac{1}{2} \ , \qquad  k = 1 \ .
\end{equation}

\subsection{Triple Bottom Line's Efficient Social Relations Portfolio}\label{SRP3Ps}

Having characterized the (graphical and mathematical) behaviors of TBL's \textit{Profit}, \textit{People}, and \textit{Planet} across each of the four sociality forms, we proceed to put together these separate pieces (step~f) and compose full analytical functions for the three pillars; namely, ${\mathcal Pr} (\mathcal C, \mathcal A, \mathcal E, \mathcal M)$ for \textit{Profit}, ${\mathcal Pe} (\mathcal C, \mathcal A, \mathcal E, \mathcal M)$ for \textit{People}, and ${\mathcal Pl} (\mathcal C, \mathcal A, \mathcal E, \mathcal M)$ for \textit{Planet}. As the corresponding mathematical parts are already normalized by construction, one may simply add them together in a weighted fashion:\footnote{Strictly speaking, \eqref{PrFull}--\eqref{PlFull} are the lowest order representations, which, for example, ignore potential higher order cross-terms. As mentioned before, this work mainly focuses on simplicity and essence of the issue; higher order terms may prove relevant in future refinements of the treatment.} 
\begin{align}
{\mathcal Pr} (\mathcal C, \mathcal A, \mathcal E, \mathcal M) &= a_{\mathcal C}\, {\mathcal Pr} (\mathcal C) + a_{\mathcal A} \, {\mathcal Pr} (\mathcal A) + a_{\mathcal E} \, {\mathcal Pr} (\mathcal E) + a_{\mathcal M} \, {\mathcal Pr} (\mathcal M)\ , \label{PrFull}\\
{\mathcal Pe} (\mathcal C, \mathcal A, \mathcal E, \mathcal M) &= b_{\mathcal C}\, {\mathcal Pe} (\mathcal C) + b_{\mathcal A} \, {\mathcal Pe} (\mathcal A) + b_{\mathcal E} \, {\mathcal Pe} (\mathcal E) + b_{\mathcal M} \, {\mathcal Pe} (\mathcal M) \ , \label{PeFull}\\
{\mathcal P\ell} (\mathcal C, \mathcal A, \mathcal E, \mathcal M) &= c_{\mathcal C}\, {\mathcal P\ell} (\mathcal C) + c_{\mathcal A} \, {\mathcal P\ell} (\mathcal A) + c_{\mathcal E} \, {\mathcal P\ell} (\mathcal E) + c_{\mathcal M} \, {\mathcal P\ell} (\mathcal M) \ , \label{PlFull}
\end{align}
where ${\mathcal Pr} (\mathcal C)$, ${\mathcal Pr} (\mathcal A)$, ${\mathcal Pr} (\mathcal E)$, ${\mathcal Pr} (\mathcal M)$ are respectively given by \eqref{PrCS}, \eqref{PrAR}, \eqref{PrEM}, \eqref{PrMP}; ${\mathcal Pe} (\mathcal C)$, ${\mathcal Pe} (\mathcal A)$, ${\mathcal Pe} (\mathcal E)$, ${\mathcal Pe} (\mathcal M)$ repectively by \eqref{PeCS}, \eqref{PeAR}, \eqref{PeEM}, \eqref{PeMP}; and ${\mathcal P\ell} (\mathcal C)$, ${\mathcal P\ell} (\mathcal A)$, ${\mathcal P\ell} (\mathcal E)$, ${\mathcal P\ell} (\mathcal M)$ respectively by \eqref{PlCS}, \eqref{PlAR}, \eqref{PlEM}, \eqref{PlMP}. 

The coefficients $a_{X}$, $b_{X}$ and $c_{X}$ $\pbrac{X \in \cbrac{\mathcal C, \mathcal A, \mathcal E, \mathcal M}}$ denote weighted average of each profile contributing to the corresponding pillar; e.g., $a_{X} = 1/4$ represents the simple average of four sociality profiles, each contributing equally to the \textit{Profit} function. At this point, for the sake of generalizability, these coefficients are left as additional free parameters of the model, the exact values of which may empirically be calibrated, although their sum within each pillar is still constrained to 1; i.e.,
\begin{equation}\label{aSum}
\sum_X a_X = \sum_X b_X = \sum_X c_X = 1 \qquad \pbrac{X \in \cbrac{\mathcal C, \mathcal A, \mathcal E, \mathcal M}} \ .
\end{equation}

In order to construct the complete analytical function of TBL, ${\mathcal P} (\mathcal C, \mathcal A, \mathcal E, \mathcal M)$, the pillar functions \eqref{PrFull}--\eqref{PlFull} should then be combined; i.e.,\footnote{In constructing \eqref{3PsFull}, it is assumed that each pillar is of equal importance to the TBL paradigm. One could also take the simple average of three pillars (i.e., assigning the weight~$1/3$ to each) in \eqref{3PsFull}; however, this will not affect the final identification of efficient SRP, and is thus immaterial to our analysis. Assigning unequal weights would, in contrast, amount to artificially altering the relative importance of pillars within the TBL paradigm. A similar observation also holds for the weight coefficients in the pillar functions \eqref{PrFull}--\eqref{PlFull}.} 
\begin{equation}\label{3PsFull}
{\mathcal P} (\mathcal C, \mathcal A, \mathcal E, \mathcal M) = {\mathcal Pr} (\mathcal C, \mathcal A, \mathcal E, \mathcal M) + {\mathcal Pe} (\mathcal C, \mathcal A, \mathcal E, \mathcal M) + {\mathcal P\ell} (\mathcal C, \mathcal A, \mathcal E, \mathcal M) \ .
\end{equation}
Furthermore, as explained towards the end of Section~\ref{quant4elem}, the universal metarelation \eqref{fundam} is to be used as a boundary condition to eliminate one of the sociality parameters in favor of the other three. Consequently, the pillar functions \eqref{PrFull}--\eqref{PlFull}, as well as the TBL function \eqref{3PsFull}, formally depend on only three of the sociality parameters as their independent variables.

\begin{figure}
\begin{center}
\includegraphics[width=.5\textwidth]{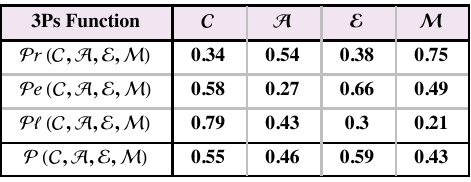}
\caption{(Illustrative) efficient SRP for each separate pillar (${\mathcal Pr}$, ${\mathcal Pe}$, and ${\mathcal P\ell}$), as well as for the entire TBL paradigm (${\mathcal P}$).}
\label{Table3Ps}
\end{center}
\end{figure}

On this account, following the procedural methodology introduced in Section~\ref{SRP}, and guided by the available literature and theoretical considerations, an analytical framework of TBL has been developed and presented as a function of sociality forms of the relational models theory. Hence, it is now possible to mathematically determine the efficient SRP (i.e., the optimal mixture of sociality parameters) separately for each pillar, a combination of only two pillars (e.g., \textit{Profit} and \textit{Planet}) as explored in some of the (older) literature (c.f., Section~\ref{LitRev}), as well as for entire TBL. To this end, computational routines (analytical and~/ or numerical) can be utilized to maximize the function under consideration (step g), after incorporating the universal metarelation \eqref{fundam} as a boundary condition to eliminate one of the sociality parameters.\footnote{For completeness, we explicitly mention that eliminating one of the sociality parameters using the boundary condition \eqref{fundam} concretely translates into, e.g., substituting $\mathcal{M}$ in all functions by $\mathcal{M} = \pbrac{1- \mathcal{C}^\gamma - \mathcal{A}^{\alpha} - \mathcal{E}^{\epsilon}}^{1/\mu}$. In addition, the optimization of each function amounts to finding its global maximum in the parameter space of sociality forms, with such substitution incorporated. The coordinate of this maximum within the parameter space (i.e., the values of sociality parameters returning function's global maximum) is then the desired efficient SRP.} The obtained values of sociality parameters maximizing each function subsequently define its corresponding efficient SRP.

As previously explained, we perform the maximization operation separately for each pillar function \eqref{PrFull}--\eqref{PlFull}, as well as for the full TBL function \eqref{3PsFull}, to facilitate a comparison of their corresponding efficient SRPs. Taking the aforementioned illustrative values of free parameters within the pillar profiles (see Appendix for a concise summary), as well as the universal metarelation's power constants \eqref{PCval}, and the simple average  $a_{X} = b_X = c_X = 1/4$, automated routines are deployed to determine the global maximum of each of those functions, and the corresponding sociality parameters' optimal values. The obtained results are summarized in Fig.~\ref{Table3Ps}. It should be emphasized that, although these efficient SRP values are, strictly-speaking, only illustrative (given the used uncalibrated free parameters), they are still useful in elucidating the introduced concepts, methodology and procedure, which themselves reside on robust (theoretical and empirical) research findings.\footnote{See also the related discussion in Section~\ref{Implic}.}

\begin{figure}
\begin{center}
\includegraphics[width=.49\textwidth]{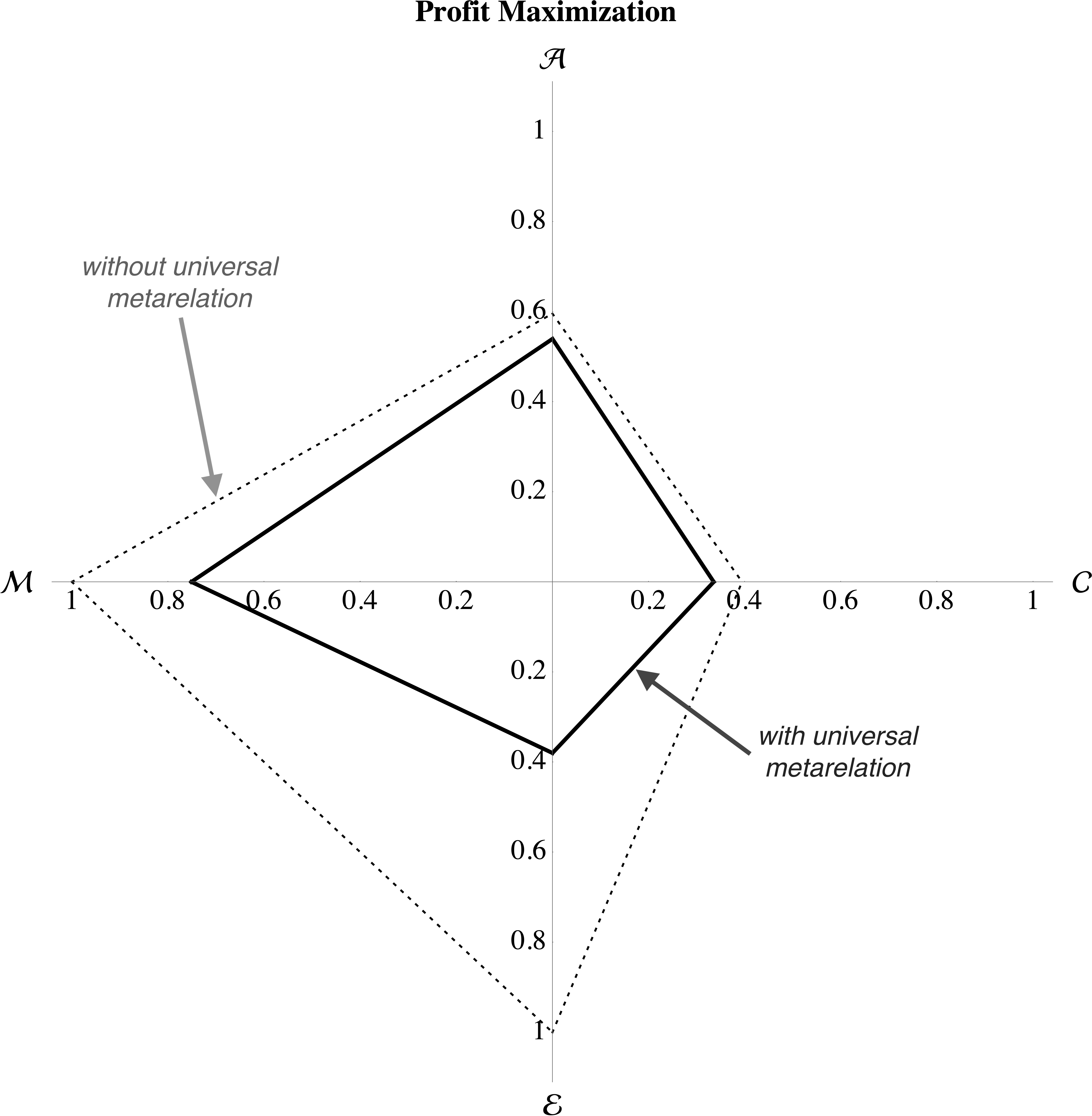}
\includegraphics[width=.49\textwidth]{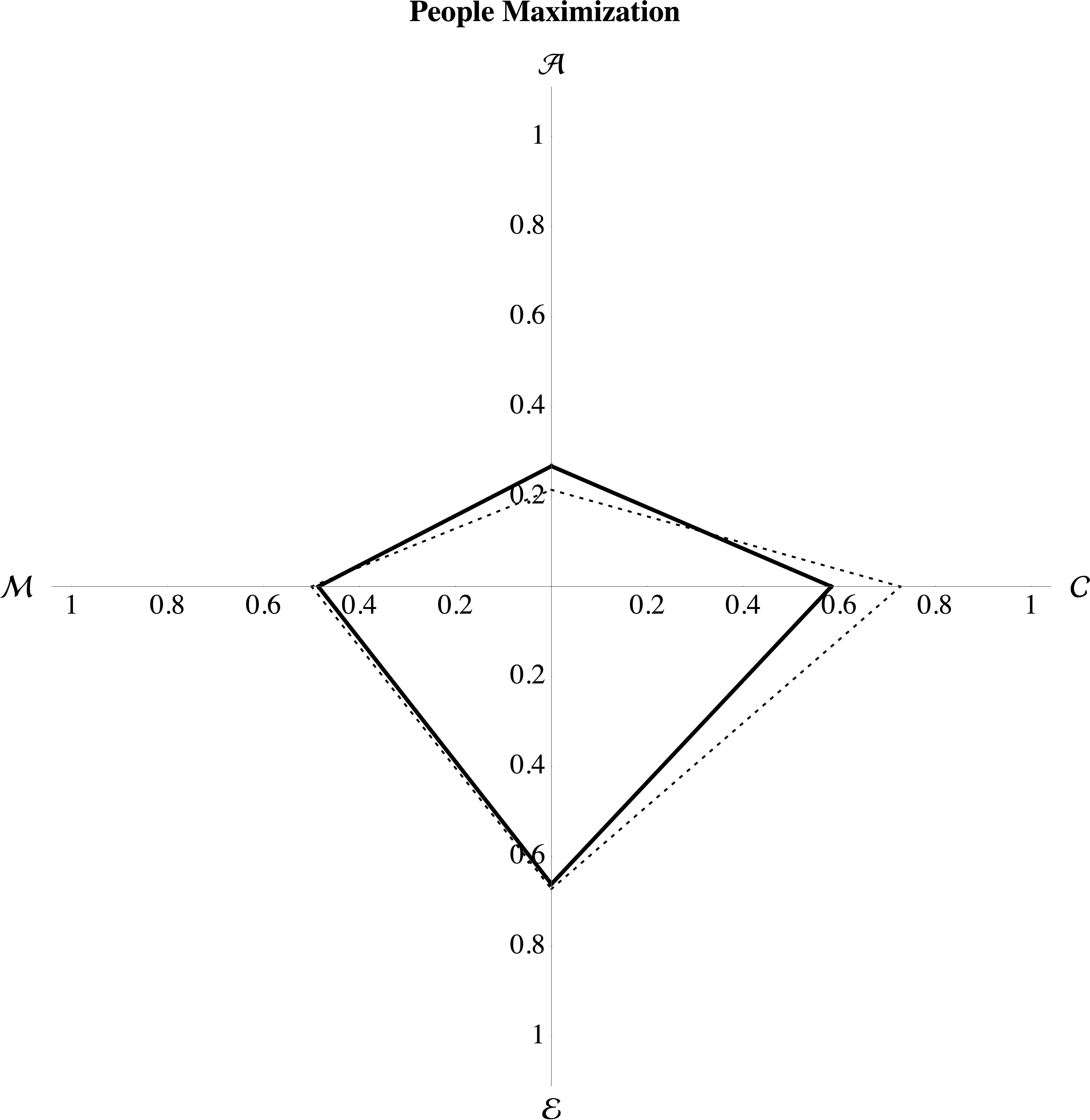}
\includegraphics[width=.49\textwidth]{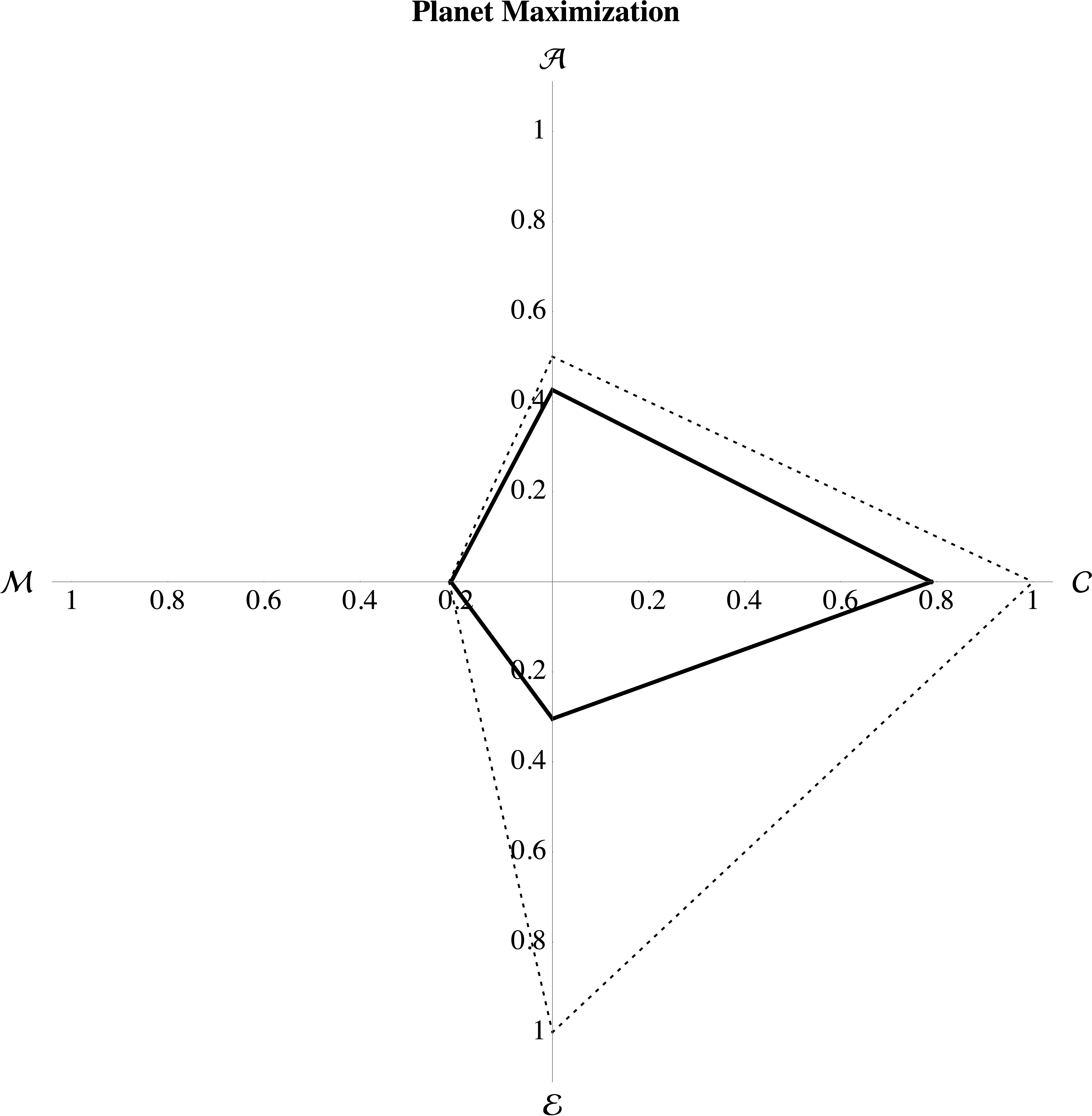}
\includegraphics[width=.49\textwidth]{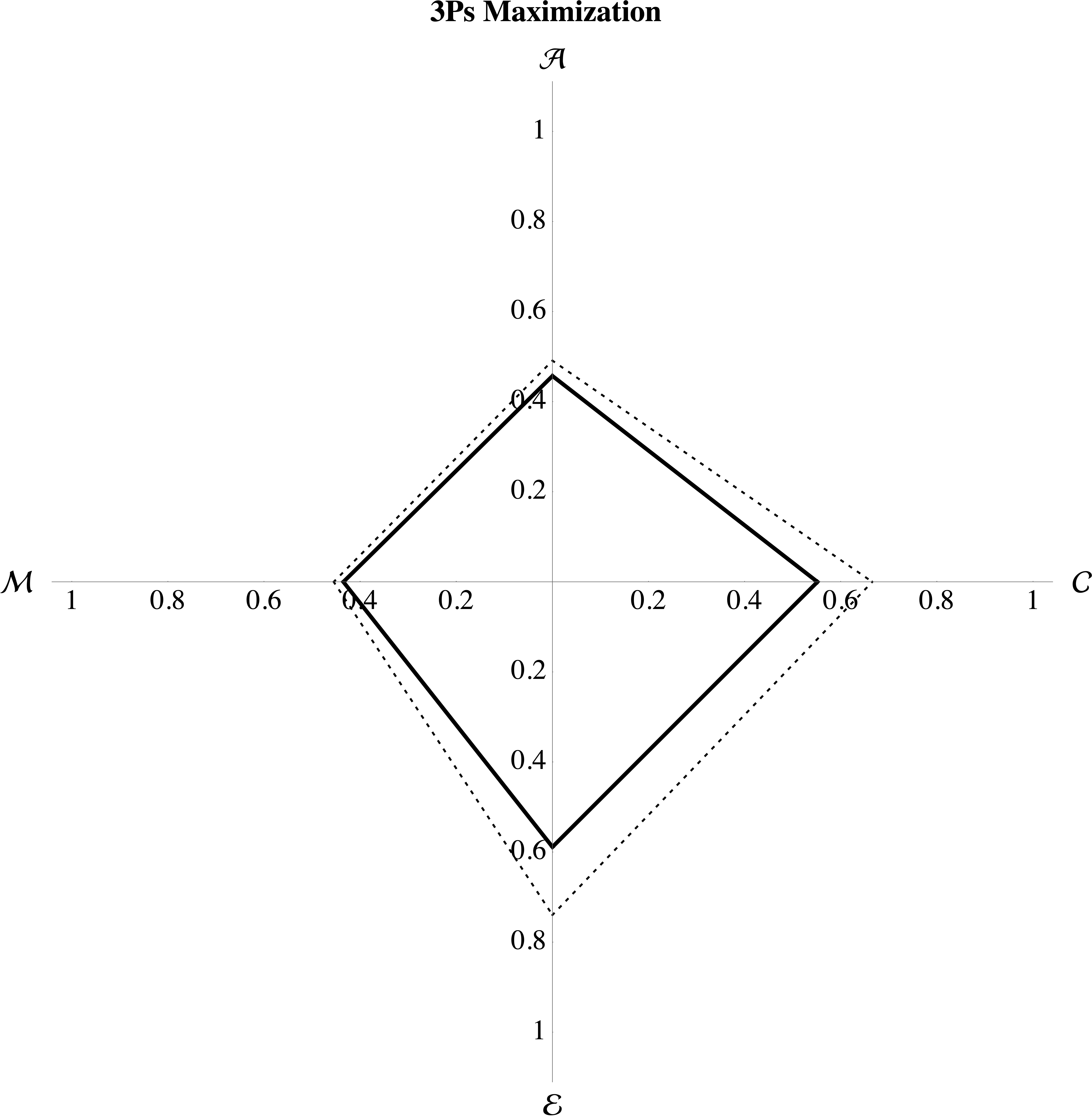}
\caption{Visual representation of the (illustrative) efficient SRP for each separate pillar, as well as for the entire TBL paradigm. Panels exhibit the optimal values of four sociality forms (solid) leading to a maximization of \textit{Profit} (top-left), \textit{People} (top-right), \textit{Planet} (bottom-left), and TBL (bottom-right), for the illustrative free parameters discussed in the text. For comparison, the corresponding values in the absence of universal metarelation \eqref{fundam} are also displayed (dotted). One notes that the existence of a universal metarelation among sociality forms leads to their internal compromises, and thus an overall moderation of the efficient SRP values to different degrees (i.e., the solid curves being mainly enclosed within the dotted curves).}
\label{Pmaxplot}
\end{center}
\end{figure}

For visual reference, the derived efficient SRP configurations are graphically displayed within the panels of Fig.~\ref{Pmaxplot} (solid curves). For comparison, the corresponding sociality parameter values in the \textit{absence} of universal metarelation \eqref{fundam} are also depicted (dotted curves); that is, in case all four sociality forms would (wrongfully) still be considered as independent within a metarelational configuration. These latter values then simply correspond to the coordinates of maxima of the curves sketched in Figs.~\ref{Prplots},~\ref{Peplots},~and~\ref{Plplots}. It is interesting to explicitly observe that the existence of a universal metarelation among sociality forms compels their internal tradeoffs to various degrees within each scenario. 

\section{Discussion and Future Research}\label{Implic}

In Section~\ref{Applic}, it was demonstrated how the analytical formulation of (meta)relational models theory and the SRP optimization methodology could in practice be utilized, by discussing their application to the TBL paradigm. In particular, the mathematical approach has enabled deriving optimal configurations of sociality forms (i.e., efficient SRPs) for actors aiming at maximizing a certain pillar or TBL as a whole. In this section, we take a closer look at the implications of obtained results, further applicability of the framework, some limitations of the study, as well as potential avenues for improvements and future research.

An examination of the \textit{Profit} panel in Fig.~\ref{Pmaxplot} (top-left) reveals that, in case the universal metarelation \eqref{fundam} is (wrongfully) not taken into account, financial profits are expected to be maximized for extreme values of MP and EM. However, given the important impacts of AR and CS on financial profits, as substantiated by research findings (c.f., Sections~\ref{PRCS} and \ref{PRAR}), one observes that the internal tension among relational models (parametrized by the universal sociality metarelation) results in an optimized configuration which is significantly different: although, as expected, MP still plays a considerable role in financial profits, the significance of EM is diminished in favor of AR. Therefore, the efficient SRP for profit is predicted to consist of dominating MP and AR sociality forms, nonetheless complemented by material contributions from EM and CS.\footnote{A domination of MP and AR within business relationships is in line with the traditional application of economic theory to the corporate environment and its classic hierarchies (i.e., transactional agency behavior and division of labor). Nevertheless, our quantitative optimization, inspired by the vast qualitative research, explicitly demonstrates importance of the remaining two models within corporate relationships primarily aimed at maximizing financial results.}

On another hand, the \textit{People} panel (top-right) exhibits a less prominent alteration of the optimization of people's well-being profile by the universal metarelation, where a moderate decrease in CS and a small increase in AR are observed, while EM and MP practically remain unaffected. In particular, the efficient SRP for people's well-being is realized when the presence of CS and EM within relationships is relatively emphasized (as expected), while that of AR is less encouraged. Interestingly, a moderate level of MP continues to contribute to the optimal well-being of people, reflecting the ever-present need for individualism, personal identity and accomplishment, which should be satisfied in addition to the other needs.

The significance of internal sociality tradeoffs within meaningful social relations (i.e., the universal metarelation) is, once more, notably affirmed for the longterm sustainability, as depicted in the \textit{Planet} panel (bottom-left). Similar to the financial profits case, it implies a considerable moderation of EM in favor of AR, whereas the most prominent role is still retained by CS, with less important MP largely unaltered. As concurred by the literature, the efficient SRP for \textit{Planet} is, thus, predicted to contain a dominant CS component (signifying collective harmony), moderate levels of AR (imposing leadership and efficacy), as well as non-negligible levels of EM and MP (implying partnership and innovation).

The combination of all three pillars in full TBL is displayed in the 3Ps panel (bottom-right). It is predicted to require a `healthy' balance of socialities at its optimum; in other words, all four sociality forms seem to contribute more or less on an equal footing to TBL's efficient SRP --- a conclusion virtually unaffected by their internal compromises (due to the universal metarelation). This is perhaps not surprising, as competing effects and interplays of the three pillars pull the sociality parameters in different directions for their own respective optimums (c.f., the other three panels of Fig.~\ref{Pmaxplot}). Hence, a simultaneous optimization of all three pillars ultimately culminates in moderate values of all sociality parameters. One, therefore, infers that the universal metarelation primarily moderates extreme values of sociality parameters within the pillars, as a result of four relational models informing and precluding one another once combined.

By incorporating all three TBL pillars in our study, we have demonstrated clear differences in the relational models most suitable for each desired pillar or outcome. Comparing pillars' efficient SRPs, the panels visually convey that, although each pillar's optimization entails a relatively unique emphasis on certain sociality forms (e.g., MP and AR for \textit{Profit}), influence of the remaining schemata is still largely non-negligible within the relationship. In practice, all four forms contribute to all pillars to various degrees, and should therefore be taken into account, which becomes manifestly evident in the full 3Ps combination. Be that as it may, it is worth emphasizing that, thanks to the introduced mathematical modeling approach, such \textit{qualitative} statements can now firmly be placed on precise \textit{quantitive} grounds, making it possible for them to be measured, monitored, reported and even actively managed by organizations.

The introduced concepts, modeling approach, methodology and procedure in this work are grounded in robust research, (mathematical) reasoning and empirical findings within the literature; nevertheless, an obvious shortcoming of the obtained numerical TBL results (c.f., Fig.~\ref{Table3Ps}) is that they are derived using the various uncalibrated values of free parameters present within the framework. Consequently, as mentioned earlier, the obtained values of these results also remain illustrative themselves; in principle, they should be checked once all free parameters are empirically calibrated by statistical data using, e.g., regression methods. To be specific, the analytical framework of (meta)relational models theory involves the four power constants $\gamma, \alpha, \epsilon, \mu$ in \eqref{fundam} as free parameters, whose values must be fixed by designing controlled experiments. Furthermore, as far as the TBL application is concerned, the mathematical profile of each pillar as a function of a sociality parameter contains four free parameters $k,l,m,n$ (c.f., Fig.~\ref{3PsTables} in Appendix). In addition, the weight coefficients $a_{X}$, $b_{X}$ and $c_{X}$, in \eqref{PrFull}, \eqref{PeFull} and \eqref{PlFull} respectively, may formally be considered as free parameters. As generally the case within (natural) sciences, the values of these parameters cannot be predicted by the current framework alone, and experimental data are necessary for calibrating them, complementing the formal description and connecting theory with observation. 

In this spirit, exploring the empirical validation, verification, and falsification of the presented framework and its conclusions merits a full-fledged experimental treatment. The road ahead for future work, among others, consists of carefully designing controlled experimental setups to quantitatively explore and measure the relative strength of sociality forms, as well as their degree of preclusion, within (dyadic) relationships. This is necessary in order to calibrate the analytical framework of (meta)relational models theory; i.e., fixing the power constants of sociality metarelation \eqref{fundam}, and verifying its universality. Regarding the discussed TBL paradigm, the quantitative connection between proxies and sociality parameters, as well as the relative importance of each profile in constructing pillars and full TBL, must experimentally be examined and determined. In this manner, the true values of framework's free parameters can empirically be calibrated by fitting the theoretical model to the collected data. Our introduced mathematical modeling approach may, hence, inspire new~/ novel experimental research directions and methods for extracting specific quantitative data in social and~/ or behavioral sciences and related disciplines.

Moreover, a comprehensive sensitivity analysis may be appropriate, in order to map out the severity of efficient SRPs' sensitivity (as dependent output variable) to various free parameters of the theory (as independent input variables), whereby the most important and relevant free parameters (whose precise input heavily affects output) can be distinguished from those less relevant (where output appears more robust against). As a result, valuable insights can be obtained into internal dynamics of the analytical framework, as well as the most prominent factors impacting the final optimal configuration outcomes. Notwithstanding, it is conceivable that, given the profiles' overall shapes and behaviors (Figs.~\ref{Prplots},~\ref{Peplots},~and~\ref{Plplots}) already motivated and justified by the empirical research literature, and their significance in determining the final outcome, the actual efficient SRPs after calibration may still (roughly) reside within the same ballpark as the quoted values in Fig.~\ref{Table3Ps}, indicating an overall robustness of the model. 

From a modelbuilding perspective, the presented simplified model may further be extended to include additional considerations, if necessitated by (new) empirical or theoretical findings, which enable more refined analyses. As an interesting example, one might consider possible (secondary) \textit{entailment} effects among sociality forms, such as a potentially positive correlation between `money' and `power' (i.e., an entailment between MP and AR). Such secondary effect could be incorporated by, e.g., including a cross-term operator $\mathcal M \times \mathcal A$ currently ignored in the model, as the latter is only concerned with \textit{preclusion} of socialities, through the universal metarelation \eqref{fundam}, as the primarily effect. In addition, complementary refinements distinguishing between positive and negative forms of relational models (e.g., positive and negative reciprocity in EM) might be considered. To this end, the parameter domain in \eqref{pardom} can be extended to $[-1,1]$, allowing for the introduction of `asymmetrical' profiles, which may exhibit different behaviors in the positive and negative regions. In this fashion, comprehensive and sophisticated `real-world' scenarios could be imported into the framework and quantitatively explored.

On the theoretical side, another relevant issue concerns actors not necessarily having similar intentions regarding their social interaction (e.g., pretense \citep{Fiske:2012}). Such misalignment of interaction objectives could then (heavily) complicate its optimization, requiring a more sophisticated approach to accommodate the enlarged space of possible realistic scenarios. Similarly, considerations regarding agency misalignment (i.e., managers' decisions primarily motivated by their own personal benefits rather than firm's interests, against their fiduciary duty) may be included within the model, among others, potentially affecting the behavior of profits as a function of MP (c.f., Section~\ref{PRMP}). Furthermore, the mathematical framework can prove a convenient tool for investigating an important question proposed earlier (c.f., Section~\ref{quant4elem}); namely, whether multiple actors’ social relations could be reducible to a collection of their dyadic relationships or should rather be considered as one collective emergent `group relationship'. Through numerical combinatoric techniques, for instance, the model can be utilized to explore such group interaction dynamics.

It is worth noting that although this work is chiefly theoretical in nature, its application landscape is anticipated to be quite large, as elaborated in the previous section. In particular, the methodology presented in Section~\ref{Funrel} is generalizable by construction, and can be applied to \textit{any} arbitrary concept involving (metarelational) combinations of sociality forms requiring optimization (c.f., Section~\ref{LitRev} for examples studied within the literature). In this spirit, the devised approach can, for instance, be employed to (mathematically) explore if, and to what extent, certain social, political, ethical, moral, or even spiritual systems and ideologies (or any other system involving meaningful social interactions of people, for that matter) are successful in fulfilling their advertised objectives (e.g., minimizing human suffering, maximizing societal benefits, maximization of political stability, etc.), by deriving their efficient SRPs (`\textit{ought}') and comparing them with the actual metarelational configurations applied by people (`\textit{is}'). To this end, mathematical profiles pertaining to a specific system as functions of the sociality parameters should be constructed, and the associated procedural steps prescribed in this treatment followed. This is admittedly a formidable undertaking, which can, nevertheless, be performed in a systematic fashion, as demonstrated in this work. Utilizing hard data, the mathematical tool is, hence, capable of (potentially conclusively) settling various disputes within social sciences, philosophy and other fields (or even starting new ones), by singling out and proving the `right' answer (or at least illuminating the correct path), in analogy with natural sciences. 

Finally, it should be emphasized that, in any case under consideration, (nature and degree of) a potential discrepancy between the in-practice applied metarelational models and the corresponding efficient SRP (i.e., `\textit{is}' vs. `\textit{ought}') serves as a solid indicator regarding the effectiveness of stakeholder and change management endeavors within and~/ or across groups and organizations. Such indicator allows for proper and necessary (corrective) actions to be undertaken, in order to steer these efforts toward the right direction in a concrete, measurable, and traceable manner.


\section{Conclusion}\label{Disc}

In this interdisciplinary treatment, we have examined the optimal nature of social interaction between two actors who intend to achieve certain mutually-agreed objectives. Inspired by the vast amount of literature attempting to address this question in a qualitative manner, we instead approached the issue as a mathematical optimization problem, and introduced modeling techniques --- ubiquitously employed in natural sciences --- to achieve the desired quantification. In doing so, we aimed at developing a generalizable framework with a broad range of applicability, capable of producing quantitative and measurable results.

In particular, we developed an analytical approach to the seminal (meta)relational models theory, which systematically categorizes all human interactions into (combinations of) four basic schemata of Communal Sharing, Authority Ranking, Equality Matching, and Market Pricing. The analytical framework allowed us to demonstrate that, once used in combination as ingredients to conjoin a specific social situation, the four sociality forms are entangled by one elementary metarelation operating at the heart of theory. This universal metarelation encodes an inherent tension among the four models, compelling them to inform and preclude one another to various degrees depending on their employed intensities within the social situation.

Elevating the developed analytical approach, a generalizable and powerful methodology was proposed, capable of mathematically determining the desired configuration of four relational models within a meaningful social interaction, which would result in the optimization of a (set of) predefined objective(s). Taking cues from financial asset~/ portfolio management in economics, we introduced the concept of efficient \textit{Social Relations Portfolio} (SRP), which corresponds to the aforementioned optimal combination of sociality forms between dyadic actors interested in optimizing certain interaction purposes. Hence, by deriving the efficient SRP for a specific contextual interaction objective, the optimization problem ubiquitously encountered within the literature is resolved. Moreover, the identification of efficient SRP enables actors to engage in effective stakeholder management; i.e., to quantitatively measure and monitor nature and composition of their social relationships, evaluate their efficacy with respect to predefined goals, and potentially undertake corrective actions in order to steer them towards optimum configurations. As a result, SRP management may prove cardinal to socially-related decision making and change management within and across firms, instigating and illuminating the necessary changes in organizational culture, way of working, and stakeholder management. The presented methodology is by construction generic in nature, and can be utilized for the optimization of any arbitrary social interaction goal involving metarelational models.

As a concrete and important application, we considered the Triple Bottom Line (TBL) paradigm, encompassing three pillars of \textit{Profit}, \textit{People}, and \textit{Planet}. Specifically, we integrated the large body of available (but predominantly unconnected) literature in economics, sociology, stakeholder management, organizational culture, and sustainability, to construct a unified understanding of behaviors of the pillars as functions of the four sociality forms. Employing mathematical modeling techniques, the graphical behaviors of pillars were cast into corresponding analytical profiles, whereby their quantitative optimization could be achieved. The optimization task was performed for each pillar separately as well as for entire TBL, and the corresponding efficient SRPs were derived for the illustrative values of framework's free parameters, unlocking important clues in the optimal composition of relational models within each pillar and within TBL as a whole (see Appendix for a concise summary). In this manner, a concrete method was presented for practitioners to engage in TBL optimization, measuring and reporting.

Furthermore, we discussed the potentially extensive application landscape of introduced framework. This included, among others, its capacity for (conclusively) settling qualitative disputes within social sciences and beyond by providing irrefutable mathematical resolutions, presenting means of exploring the nature of social interactions among many actors, as well as accommodating objective misalignments akin to pretense and agency behavior. Several shortcomings and possible extensions of the simplified model were also scrutinized, in addition to inspirations for future theoretical and experimental research.

Finally, our presented generalizable approach, interdisciplinary methods, and overall analytical rigor may pave the way for enabling and encouraging mathematical modelbuilding efforts in social and behavioral sciences, currently underrepresented in these disciplines. Such novel tools could prove invaluable, by setting the theoretical foundations of these important fields on strong analytical grounds, enhancing their (quantitative) predictive power, as well as opening up avenues to previously out of reach research opportunities, for better understanding of the fascinating nature of human behavior and social relations.

\section*{Acknowledgment}

We are grateful to Alan P. Fiske for reading the manuscript and providing valuable comments, improving its overall quality and clarity. A.F. thanks Viktor Tchistiakov for interesting suggestions, and Jack van der Veen for early discussions.

\appendix

\section*{Appendix}\label{Appendix}
\addcontentsline{toc}{section}{Appendix}

In this appendix, a summary of constructed mathematical profiles for the three pillars of TBL (\textit{Profit}, \textit{People}, and \textit{Planet}) as functions of the sociality forms (CS, AR, EM, and MP) is provided for reference and convenience, along with their corresponding constraints and illustrative values discussed in the text. 

The sociality forms are parametrized by \eqref{pardom},
\begin{equation*}
\mathcal{C} \in [0,1] \ , \qquad  \mathcal{A} \in [0,1] \ , \qquad  \mathcal{E} \in [0,1] \ , \qquad  \mathcal{M} \in [0,1] \ , 
\end{equation*}
which are interrelated by the elementary and universal sociality metarelation \eqref{fundam},
\begin{equation*}
\mathcal{C}^\gamma + \mathcal{A}^{\alpha} + \mathcal{E}^{\epsilon} + \mathcal{M}^{\mu} = 1 \ .
\end{equation*}
The power constants in this metarelation obey the inequality \eqref{powerconst},
\begin{equation*}
0 < \gamma < \alpha < \epsilon < \mu \ ,
\end{equation*}
for which, throughout this treatment, the illustrative values \eqref{PCval} are employed,
\begin{equation*}
\gamma = 1 \ , \quad \alpha = 2 \ , \quad \epsilon = 3 \ , \quad \mu = 4 \ .
\end{equation*}

\begin{figure}
\begin{center}
\includegraphics[width=.7\textwidth]{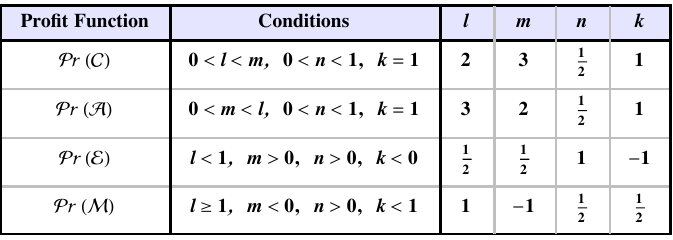}
\includegraphics[width=.7\textwidth]{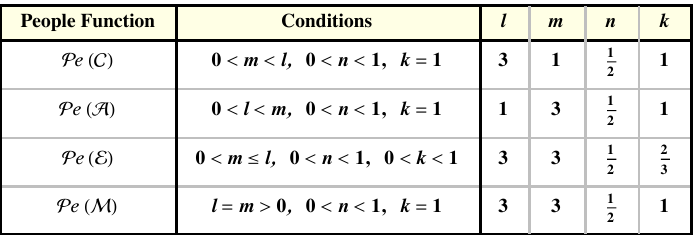}
\includegraphics[width=.7\textwidth]{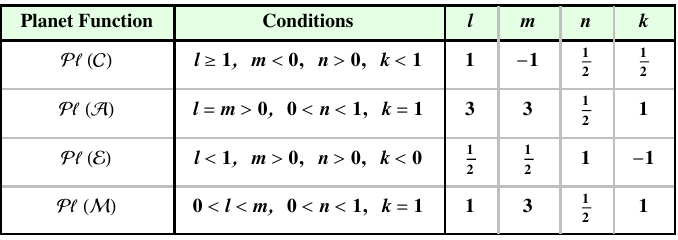}
\caption{Summary of the imposed constraints and variables' illustrative values for analytical profiles of the three TBL pillars (${\mathcal Pr}$, ${\mathcal Pe}$, and ${\mathcal P\ell}$) as a function of sociality parameters (${\mathcal C}$, ${\mathcal A}$, ${\mathcal E}$, and ${\mathcal M}$), discussed in this treatment.}
\label{3PsTables}
\end{center}
\end{figure}

Interestingly, all pillars' analytical profiles may be obtained by appropriately adjusting the four variables ($k,l,m,n$) within a single normalized function \eqref{ParGen}, involving a power part and an exponential part:
\begin{equation*}
P (X) = \frac{1}{N} \cbrac{ X^{l} \times \exp \tbrac{-m \pbrac{\frac{X}{1 - k \, X}}^{n}} } \qquad \quad  \pbrac{ X \in \cbrac{\mathcal{C}, \mathcal{A}, \mathcal{E}, \mathcal{M}} } \ ,
\end{equation*}
\begin{equation*}
N \equiv \int_{0}^{1} X^{l} \exp \tbrac{-m \pbrac{\frac{X}{1 - k \, X}}^{n}}  dX \ .
\end{equation*}
The normalization of this function facilitates a relative comparison of the different profile functions, as well as the possibility of their addition.

The conditions for constructing the analytical profile of each of the TBL pillars per sociality parameter, as well as variables' illustrative values utilized within this treatment, are summarized in Fig.~\ref{3PsTables}. The corresponding profile curves are plotted within panels of Figs.~\ref{Prplots},~\ref{Peplots},~and~\ref{Plplots}. The derived (illustrative) SRPs are given in Figs.~\ref{Table3Ps}~and~\ref{Pmaxplot}.


\begin{thebibliography}{99}
\addcontentsline{toc}{section}{References}


\bibitem[Adams, 1963]{Adams:1963}
 Adams, S. J. (1963). Towards an understanding of inequity. \textit{Journal of Abnormal and Social Psychology, 67}(5), 422.
 
 \bibitem[Ahuvia, 2002]{Ahuvia:2002}
Ahuvia, A. C. (2002). Individualism/collectivism and cultures of happiness: A theoretical conjecture on the relationship between consumption, culture and subjective well-being at the national level. \textit{Journal of Happiness Studies, 3}(1), 23--36.

\bibitem[Albert, 1992]{Albert:1992}
Albert, M. (1992). The Rhine model of capitalism: an investigation. \textit{European Business Journal, 4}(3), 8.

\bibitem[Albert, 1993]{Albert:1993}
Albert, M. (1993). \textit{Capitalism against capitalism} (Vol. 1). John Wiley \& Sons Incorporated.

\bibitem[Alhaddi, 2015]{Alhaddi:2015}
Alhaddi, H. (2015). Triple bottom line and sustainability: A literature review. \textit{Business and Management Studies, 1}(2), 6--10.

\bibitem[Ames \& Flynn, 2007]{Ames:2007}
Ames, D. R., \& Flynn, F. J. (2007). What breaks a leader: The curvilinear relation between assertiveness and leadership. \textit{Journal of personality and social psychology, 92}(2), 307.

\bibitem[Amnesty International, 2018]{Amnesty:2018}
Amnesty International (2018). Global Death Sentences and Executions: Report, Death Sentences and Executions 2018. \textit{Amnesty International} (\url{https://www.amnesty.org/download/Documents/ACT5098702019ENGLISH.PDF}).

\bibitem[Angus-Leppan et al., 2010]{Angus:2010}
Angus-Leppan, T., Metcalf, L., \& Benn, S. (2010). Leadership styles and CSR practice: An examination of sensemaking, institutional drivers and CSR leadership. \textit{Journal of Business Ethics, 93}(2), 189--213.

\bibitem[Arendt et al., 2022]{Arendt:2022}
Arendt, J. F., Kugler, K. G., \& Brodbeck, F. C. (2022). Being on the same page about social rules and norms: Effects of shared relational models on cooperation in work teams. \textit{Group Processes \& Intergroup Relations}, 13684302221088506.

\bibitem[Bakker \& Demerouti, 2007]{Bakker:2007}
Bakker, A. B., \& Demerouti, E. (2007). The job demands-resources model: State of the art. \textit{Journal of managerial psychology, 22}(3), 309--328.

\bibitem[Beehr, 1998]{Beehr:1998}
Beehr, T. (1998). An organizational psychology meta-model of occupational stress. \textit{Theories of organizational stress}, 6-27.

\bibitem[Bendell \& Little, 2015]{Bendell:2015}
Bendell, J., \& Little, R. (2015). Seeking sustainability leadership. \textit{Journal of Corporate Citizenship, }(60), 13--26.

\bibitem[Blau, 1964]{Blau:1964}
Blau, P. M. (1964). Justice in social exchange. \textit{Sociological Inquiry, 34}(2), 193--206.

\bibitem[Blomme et al., 2014]{Blomme:2014}
Blomme, R. J., van der Veen, J. A., \& Venugopal, V. (2014). Silver Lining of a Dark Cloud: Using social innovation to make the supply chain a crisis-buster. \textit{Challenging Organisations and Society 3}(2), 544--560.

\bibitem[Bodla et al., 2019]{Bodla:2019}
Bodla, A. A., Tang, N., Van Dick, R., \& Mir, U. R. (2019). Authoritarian leadership, organizational citizenship behavior, and organizational deviance. \textit{Leadership \& Organization Development Journal}.

\bibitem[Boer et al., 2011]{Boer:2011}
Boer, N. I., Berends, H., \& Van Baalen, P. (2011). Relational models for knowledge sharing behavior. \textit{European management journal, 29}(2), 85--97.

\bibitem[Bolender, 2015]{Bolender:2015}
Bolender, J. (2015). Digital social mind. Andrews UK Limited.

\bibitem[Bolino et al., 2013]{Bolino:2013}
Bolino, M. C., Klotz, A. C., Turnley, W. H., \& Harvey, J. (2013). Exploring the dark side of organizational citizenship behavior. \textit{Journal of Organizational Behavior, 34}(4), 542--559.

\bibitem[Brandts et al., 2008]{Brandts:2008}
Brandts, J., Riedl, A., \& Van Winden, F. (2008). On competition and well-being. An experimental investigation into rivalry, social disposition, and subjective well-being.

\bibitem[Bridoux \& Stoelhorst, 2016]{Bridoux:2016}
Bridoux, F., \& Stoelhorst, J. W. (2016). Stakeholder relationships and social welfare: A behavioral theory of contributions to joint value creation. \textit{Academy of Management Review, 41}(2), 229--251.

\bibitem[Burns et al., 2015]{Burns:2015}
Burns, H., Diamond-Vaught, H., \& Bauman, C. (2015). Leadership for sustainability: Theoretical foundations and pedagogical practices that foster change. \textit{International Journal of Leadership Studies}.

\bibitem[Carter \& Rogers, 2008]{Carter:2008}
Carter, C. R., \& Rogers, D. S. (2008). A framework of sustainable supply chain management: moving toward new theory. \textit{International journal of physical distribution \& logistics management}.

\bibitem[Chuang \& Liao, 2010]{Chuang:2010}
Chuang, C. H., \& Liao, H. U. I. (2010). Strategic human resource management in service context: Taking care of business by taking care of employees and customers. \textit{Personnel psychology, 63}(1), 153--196.

\bibitem[Cornelissen et al., 2014]{Cornelissen:2014}
Cornelissen, T., Heywood, J. S., \& Jirjahn, U. (2014). Reciprocity and Profit Sharing: Is There an Inverse U-shaped Relationship? \textit{Journal of Labor Research, 35}(2), 205--225.

\bibitem[De Cremer, 2006]{DeCremer:2006}
De Cremer, D. (2006). Affective and motivational consequences of leader self-sacrifice: The moderating effect of autocratic leadership. \textit{The Leadership Quarterly, 17}(1), 79--93.

\bibitem[De Hoogh \& Den Hartog, 2009]{DeHoogh:2009}
De Hoogh, A. H., \& Den Hartog, D. N. (2009). Neuroticism and locus of control as moderators of the relationships of charismatic and autocratic leadership with burnout. \textit{Journal of Applied Psychology, 94}(4), 1058.

\bibitem[De Hoogh et al., 2015]{DeHoogh:2015}
De Hoogh, A. H., Greer, L. L., \& Den Hartog, D. N. (2015). Diabolical dictators or capable commanders? An investigation of the differential effects of autocratic leadership on team performance. \textit{The Leadership Quarterly, 26}(5), 687--701.

\bibitem[De Witte et al., 2016]{DeWitte:2016}
De Witte, H., Pienaar, J., \& De Cuyper, N. (2016). Review of 30 years of longitudinal studies on the association between job insecurity and health and well‐being: is there causal evidence?. \textit{Australian Psychologist, 51}(1), 18--31.

\bibitem[Deutsch, 1975]{Deutsch:1975}
Deutsch, M. (1975). Equity, equality, and need: What determines which value will be used as the basis of distributive justice?. \textit{Journal of Social issues, 31}(3), 137--149.

\bibitem[Diener et al., 2009]{Diener:2009}
Diener, E., Diener, M., \& Diener, C. (2009). Factors predicting the subjective well-being of nations. In \textit{Culture and well-being} (pp. 43--70). Springer, Dordrecht.

\bibitem[Elkington, 1998]{Elkington:1998}
Elkington, J. (1998). Partnerships from cannibals with forks: The triple bottom line of 21st‐century business. \textit{Environmental quality management, 8}(1), 37--51.

\bibitem[Elkington, 2018]{Elkington:2018}
Elkington, J. (2018). 25 Years Ago I Coined the Phrase ``Triple Bottom Line.'' Here's Why It's Time to Rethink It. \textit{Harvard Business Review}, 25 June 2018.

\bibitem[Ellemers et al., 2004]{Ellemers:2004}
Ellemers, N., De Gilder, D., \& Haslam, S. A. (2004). Motivating individuals and groups at work: A social identity perspective on leadership and group performance. \textit{Academy of Management review, 29}(3), 459--478.

\bibitem[Evans \& Davis, 2005]{Evans:2005}
Evans, W. R., \& Davis, W. D. (2005). High-Performance Work Systems and Organizational Performance: The Mediating Role of Internal Social Structure. \textit{Journal of Management, 31}(5), 758--775.

\bibitem[Farzinnia, 2019]{Farzinnia:2019}
Farzinnia, A. (2019). Relational Models Theory \& UN Sustainable Development Goals. \textit{Master of Business Administration Thesis}. Amsterdam Business School, University of Amsterdam.

\bibitem[Favre \& Sornette, 2013]{Favre:2013}
Favre, M., \& Sornette, D. (2013). Categorization of exchange fluxes explains the four relational models \textit{(No. arXiv: 1312.4644)}. Swiss Finance Inst..

\bibitem[Favre \& Sornette, 2015]{Favre:2015}
Favre, M., \& Sornette, D. (2015). A generic model of dyadic social relationships. \textit{PloS one, 10}(3), e0120882.

\bibitem[Fehr \& G\"achter, 2002]{Fehr:2002}
Fehr, E., \& G\"achter, S. (2002). Altruistic punishment in humans. \textit{Nature, 415}(6868), 137.

\bibitem[Fiske, 1991]{Fiske:1991}
Fiske, A. P. (1991). Structures of social life: The four elementary forms of human relations. New York: Free Press.

\bibitem[Fiske, 1992]{Fiske:1992}
Fiske, A. P. (1992). The four elementary forms of sociality: framework for a unified theory of social relations. \textit{Psychological review, 99}(4), 689.

\bibitem[Fiske, 2004]{Fiske:2004}
Fiske, A. P. (2004). Relational models theory 2.0. In N. Haslam (Ed.), \textit{Relational models theory: A contemporary overview:} 3--25. Mahwah, NJ: Lawrence Erlbaum Associates.

\bibitem[Fiske, 2012]{Fiske:2012}
Fiske, A. P. (2012). Metarelational models: Configurations of social relationships. \textit{European Journal of Social Psychology, 42}(1), 2--18.

\bibitem[Fiske \& Haslam, 2005]{Fiske:2005}
Fiske, A. P., \& Haslam, N. (2005). The four basic social bonds: Structures for coordinating interactions. In M. W. Baldwin (Ed.), \textit{Interpersonal cognition:} 267--298. New York: Guilford Press.

\bibitem[Franken \& Brown, 1995]{Franken:1995}
Franken, R. E., \& Brown, D. J. (1995). Why do people like competition? The motivation for winning, putting forth effort, improving one's performance, performing well, being instrumental, and expressing forceful/aggressive behavior. \textit{Personality and individual differences, 19}(2), 175--184.

\bibitem[Gaudet \& Tremblay, 2017]{Gaudet:2017}
Gaudet, M. C., \& Tremblay, M. (2017). Initiating structure leadership and employee behaviors: The role of perceived organizational support, affective commitment and leader–member exchange. \textit{European Management Journal, 35}(5), 663--675.

\bibitem[Giessner \& Van Quaquebeke, 2010]{Giessner:2010}
Giessner, S., \& Van Quaquebeke, N. (2010). Using a relational models perspective to understand normatively appropriate conduct in ethical leadership. \textit{Journal of Business Ethics, 95}(1), 43--55.

\bibitem[Gittell \& Douglass, 2012]{Gittell:2012}
Gittell, J. H., \& Douglass, A. (2012). Relational bureaucracy: Structuring reciprocal relationships into roles. \textit{Academy of management review, 37}(4), 709--733.

\bibitem[Global Peace Index, 2019]{GPI:2019}
Global Peace Index: Index, G. P. (2019). Measuring the state of global peace. \textit{Institute for Economics \& Peace} (\url{https://visionofhumanity.org/app/uploads/2019/07/GPI-2019web.pdf}).

\bibitem[Goel, 2010]{Goel:2010}
Goel, P. (2010). Triple Bottom Line Reporting: An Analytical Approach for Corporate Sustainability. \textit{Journal of Finance, Accounting \& Management, 1}(1).

\bibitem[Go{\'n}cz et al., 2007]{Goncz:2007}
Go{\'n}cz, E., Skirke, U., Kleizen, H., \& Barber, M. (2007). Increasing the rate of sustainable change: a call for a redefinition of the concept and the model for its implementation. \textit{Journal of Cleaner Production, 15}(6), 525--537.

\bibitem[Grant, 2007]{Grant:2007}
Grant, A. M. (2007). Relational job design and the motivation to make a prosocial difference. \textit{Academy of management review, 32}(2), 393--417.

\bibitem[Gu et al., 2019]{Gu:2019}
Gu, Q., Hempel, P. S., \& Yu, M. (2019). Tough Love and Creativity: How Authoritarian Leadership Tempered by Benevolence or Morality Influences Employee Creativity. \textit{British Journal of Management}.

\bibitem[Hargreaves \& Fink, 2004]{Hargreaves:2004}
Hargreaves, A., \& Fink, D. (2004). The seven principles of sustainable leadership. \textit{Educational leadership, 61}(7), 8--13.

\bibitem[Harms et al., 2017]{Harms:2017}
Harms, P. D., Cred\'e, M., Tynan, M., Leon, M., \& Jeung, W. (2017). Leadership and stress: A meta-analytic review. \textit{The leadership quarterly, 28}(1), 178--194.

\bibitem[Harris \& Kacmar, 2006]{Harris:2006}
Harris, K. J., \& Kacmar, K. M. (2006). Too much of a good thing: The curvilinear effect of leader-member exchange on stress. \textit{The Journal of social psychology, 146}(1), 65--84.

\bibitem[Haslam \& Ellemers, 2005]{Haslam:2005}
Haslam, S. A., \& Ellemers, N. (2005). Social identity in industrial and organizational psychology: Concepts, controversies and contributions. \textit{International review of industrial and organizational psychology, 20}(1), 39--118.

\bibitem[Holtz \& Harold, 2013]{Holtz:2013}
Holtz, B. C., \& Harold, C. M. (2013). Effects of leadership consideration and structure on employee perceptions of justice and counterproductive work behavior. \textit{Journal of Organizational Behavior, 34}(4), 492--519.

\bibitem[Hussain et al., 2018]{Hussain:2018}
Hussain, N., Rigoni, U., \& Orij, R. P. (2018). Corporate governance and sustainability performance: Analysis of triple bottom line performance. \textit{Journal of business ethics, 149}, 411--432.

\bibitem[Jiang et al., 2017]{Jiang:2017}
Jiang, W., Zhao, X., \& Ni, J. (2017). The impact of transformational leadership on employee sustainable performance: The mediating role of organizational citizenship behavior. \textit{Sustainability, 9}(9), 1567.

\bibitem[Jones, 1995]{Jones:1995}
Jones, T. M. (1995). Instrumental stakeholder theory: A synthesis of ethics and economics. \textit{Academy of management review, 20}(2), 404--437.

\bibitem[Jones, 2010]{Jones:2010}
Jones, D. A. (2010). Does serving the community also serve the company? Using organizational identification and social exchange theories to understand employee responses to a volunteerism programme. \textit{Journal of Occupational and Organizational Psychology, 83}(4), 857--878.

\bibitem[Jones et al., 2007]{Jones:2007}
Jones, T. M., Felps, W., \& Bigley, G. A. (2007). Ethical theory and stakeholder-related decisions: The role of stakeholder culture. \textit{Academy of Management Review, 32}(1), 137--155.

\bibitem[Judge et al., 2004]{Judge:2004}
Judge, T. A., Piccolo, R. F., \& Ilies, R. (2004). The forgotten ones? The validity of consideration and initiating structure in leadership research. \textit{Journal of applied psychology, 89}(1), 36.

\bibitem[Keck et al., 2018]{Keck:2018}
Keck, N., Giessner, S. R., Van Quaquebeke, N., \& Kruijff, E. (2018). When do Followers Perceive Their Leaders as Ethical? A Relational Models Perspective of Normatively Appropriate Conduct. \textit{Journal of Business Ethics}, 1--17.

\bibitem[Konovsky \& Pugh, 1994]{Konovsky:1994}
Konovsky, M. A., \& Pugh, S. D. (1994). Citizenship behavior and social exchange. \textit{Academy of management journal, 37}(3), 656--669.

\bibitem[Lambert et al., 2012]{Lambert:2012}
Lambert, L. S., Tepper, B. J., Carr, J. C., Holt, D. T., \& Barelka, A. J. (2012). Forgotten but not gone: An examination of fit between leader consideration and initiating structure needed and received. \textit{Journal of Applied Psychology, 97}(5), 913.

\bibitem[Lee et al., 2017]{Lee:2017}
Lee, S., Cheong, M., Kim, M., \& Yun, S. (2017). Never too much? The curvilinear relationship between empowering leadership and task performance. \textit{Group \& Organization Management, 42}(1), 11--38.

\bibitem[Lee \& Pinker, 2010]{Lee:2010}
Lee, J. J., \& Pinker, S. (2010). Rationales for indirect speech: the theory of the strategic speaker. \textit{Psychological review, 117}(3), 785.

\bibitem[Lejeune \& Yakova, 2005]{Lejeune:2005}
Lejeune, M. A., \& Yakova, N. (2005). On characterizing the 4 C's in supply chain management. \textit{Journal of operations Management, 23}(1), 81--100.

\bibitem[Li et al., 2018]{Li:2018}
Li, G., Liu, H., \& Luo, Y. (2018). Directive versus participative leadership: Dispositional antecedents and team consequences. \textit{Journal of Occupational and Organizational Psychology, 91}(3), 645--664.

\bibitem[Markowitz, 1952]{Markowitz:1952}
Markowitz, H. (1952). Portfolio Selection. \textit{Journal  of Finance 7}, 77--91.

\bibitem[M\"akikangas et al., 2016]{Makikangas:2016}
M\"akikangas, A., Kinnunen, U., Feldt, T., \& Schaufeli, W. (2016). The longitudinal development of employee well-being: A systematic review. \textit{Work \& Stress, 30}(1), 46--70.

\bibitem[Meglino \& Korsgaard, 2004]{Meglino:2004}
Meglino, B. M., \& Korsgaard, A. (2004). Considering rational self-interest as a disposition: organizational implications of other orientation. \textit{Journal of Applied Psychology, 89}(6), 946.

\bibitem[Moliner et al., 2013]{Moliner:2013}
Moliner, C., Mart\'inez‐Tur, V., Peir\'o, J. M., Ramos, J., \& Cropanzano, R. (2013). Perceived reciprocity and well‐being at work in non‐professional employees: Fairness or self‐interest?. \textit{Stress and Health, 29}(1), 31--39.

\bibitem[Morin et al., 2013]{Morin:2013}
Morin, A. J., Vandenberghe, C., Turmel, M. J., Madore, I., \& Maiano, C. (2013). Probing into commitment's nonlinear relationships to work outcomes. \textit{Journal of Managerial Psychology, 28}(2), 202--223.

\bibitem[Mossholder et al., 2011]{Mossholder:2011}
Mossholder, K. W., Richardson, H. A., \& Settoon, R. P. (2011). Human resource systems and helping in organizations: A relational perspective. \textit{Academy of Management Review, 36}(1), 33--52.

\bibitem[Munyon et al., 2010]{Munyon:2010}
Munyon, T. P., Hochwarter, W. A., Perrew\'e, P. L., \& Ferris, G. R. (2010). Optimism and the nonlinear citizenship behavior--Job satisfaction relationship in three studies. \textit{Journal of Management, 36}(6), 1505--1528.

\bibitem[Nielsen et al., 2017]{Nielsen:2017}
Nielsen, K., Nielsen, M. B., Ogbonnaya, C., K\"ans\"al\"a, M., Saari, E., \& Isaksson, K. (2017). Workplace resources to improve both employee well-being and performance: A systematic review and meta-analysis. \textit{Work \& Stress, 31}(2), 101--120.

\bibitem[Nowak, 2006]{Nowak:2006}
Nowak, M. A. (2006). Five rules for the evolution of cooperation. \textit{science, 314}(5805), 1560--1563.

\bibitem[Paill\'e et al., 2013]{Paille:2013}
Paill\'e, P., Boiral, O., \& Chen, Y. (2013). Linking environmental management practices and organizational citizenship behaviour for the environment: a social exchange perspective. \textit{The International Journal of Human Resource Management, 24}(18), 3552--3575.

\bibitem[Parboteeah et al., 2012]{Parboteeah:2012}
Parboteeah, K. P., Addae, H. M., \& Cullen, J. B. (2012). Propensity to support sustainability initiatives: A cross-national model. \textit{Journal of business ethics, 105}(3), 403--413.

\bibitem[Podsakoff et al., 2009]{Podsakoff:2009}
Podsakoff, N. P., Whiting, S. W., Podsakoff, P. M., \& Blume, B. D. (2009). Individual-and organizational-level consequences of organizational citizenship behaviors: A meta-analysis. \textit{Journal of applied Psychology, 94}(1), 122.

\bibitem[Post, 2005]{Post:2005}
Post, S. G. (2005). Altruism, happiness, and health: It’s good to be good. \textit{International journal of behavioral medicine, 12}(2), 66--77.

\bibitem[Quinn \& Dalton, 2009]{Quinn:2009}
Quinn, L., \& Dalton, M. (2009). Leading for sustainability: implementing the tasks of leadership. \textit{Corporate Governance: The international journal of business in society, 9}(1), 21--38.

\bibitem[Rai \& Fiske, 2011]{Rai:2011}
Rai, T. S., \& Fiske, A. P. (2011). Moral psychology is relationship regulation: moral motives for unity, hierarchy, equality, and proportionality. \textit{Psychological review, 118}(1), 57.

\bibitem[Seltzer \& Numerof, 1988]{Seltzer:1988}
Seltzer, J., \& Numerof, R. E. (1988). Supervisory leadership and subordinate burnout. \textit{Academy of management Journal, 31}(2), 439--446.

\bibitem[Seyfang, 2005]{Seyfang:2005}
Seyfang, G. (2005). Shopping for sustainability: can sustainable consumption promote ecological citizenship?. \textit{Environmental politics, 14}(2), 290--306.

\bibitem[Sheppard \& Tuchinsky, 1996]{Sheppard:1996}
Sheppard, B. H., \& Tuchinsky, M. (1996). Interfirm relationships: A grammar of pairs. \textit{RESEARCH IN ORGANIZATIONAL BEHAVIOR, VOL 18, 1996, 18}, 331--373.

\bibitem[Siegrist, 2015]{Siegrist:2015}
Siegrist, J. (2005). Social reciprocity and health: new scientific evidence and policy implications. \textit{Psychoneuroendocrinology, 30}(10), 1033--1038.

\bibitem[Sikdar, 2003]{Sikdar:2003}
Sikdar, S. K. (2003). Sustainable development and sustainability metrics. \textit{AIChE journal, 49}(8), 1928--1932.

\bibitem[Simpson et al., 2016]{Simpson:2016}
Simpson, A., Laham, S. M., \& Fiske, A. P. (2016). Wrongness in different relationships: Relational context effects on moral judgment. \textit{The Journal of social psychology, 156}(6), 594--609.

\bibitem[Slaper \& Hall, 2011]{Slaper:2011}
Slaper, T. F., \& Hall, T. J. (2011). The triple bottom line: What is it and how does it work. \textit{Indiana business review, 86}(1), 4--8.

\bibitem[Solow, 2014]{Solow:2014}
Solow, R. (2014). An almost practical step toward sustainability. In \textit{An Almost Practical Step Toward Sustainability} (pp. 11--28). RFF Press.

\bibitem[Stofberg et al., 2021]{Stofberg:2021}
Stofberg, N., Bridoux, F., Ciulli, F., Pisani, N., Kolk, A., \& Vock, M. (2021). A relational‐models view to explain peer‐to‐peer sharing. \textit{Journal of Management Studies, 58}(4), 1033--1069.

\bibitem[Stoker et al., 2001]{Stoker:2001}
Stoker, J. I., Looise, J. C., Fisscher, O. A. M., \& Jong, R. D. (2001). Leadership and innovation: relations between leadership, individual characteristics and the functioning of R\&D teams. \textit{International Journal of Human Resource Management, 12}(7), 1141--1151.

\bibitem[Suh \& Oishi, 2002]{Suh:2002}
Suh, E. M., \& Oishi, S. (2002). Subjective well-being across cultures. \textit{Online readings in psychology and culture, 10}(1), 1.

\bibitem[Tata \& Prasad, 2015]{Tata:2015}
Tata, J., \& Prasad, S. (2015). National cultural values, sustainability beliefs, and organizational initiatives. \textit{Cross Cultural Management, 22}(2), 278--296.

\bibitem[Tetrick \& LaRocco, 1987]{Tetrick:1987}
Tetrick, L. E., \& LaRocco, J. M. (1987). Understanding, prediction, and control as moderators of the relationships between perceived stress, satisfaction, and psychological well-being. \textit{Journal of Applied psychology, 72}(4), 538.

\bibitem[Thompson \& Prottas, 2006]{Thompson:2006}
Thompson, C. A., \& Prottas, D. J. (2006). Relationships among organizational family support, job autonomy, perceived control, and employee well-being. \textit{Journal of occupational health psychology, 11}(1), 100.

\bibitem[Tobin, 1958]{Tobin:1958}
Tobin, J. (1958). Liquidity Preference as Behavior Toward Risk. \textit{Review of Economic Studies 25}, 65--86.

\bibitem[Trivers, 1971]{Trivers:1971}
Trivers, R. L. (1971). The evolution of reciprocal altruism. \textit{The Quarterly review of biology, 46}(1), 35--57.

\bibitem[Tseng et al., 2020]{Tseng:2020}
Tseng, M. L., Chang, C. H., Lin, C. W. R., Wu, K. J., Chen, Q., Xia, L., \& Xue, B. (2020). Future trends and guidance for the triple bottom line and sustainability: A data driven bibliometric analysis. \textit{Environmental Science and Pollution Research, 27}, 33543-33567.

\bibitem[Tsutsumi \& Kawakami, 2004]{Tsutsumi:2004}
Tsutsumi, A., \& Kawakami, N. (2004). A review of empirical studies on the model of effort–reward imbalance at work: reducing occupational stress by implementing a new theory. \textit{Social science \& medicine, 59}(11), 2335--2359.

\bibitem[Tyler, 2006]{Tyler:2006}
Tyler, T. R. (2006). Psychological perspectives on legitimacy and legitimation. \textit{Annu. Rev. Psychol., 57}, 375--400.

\bibitem[Vecina \& Fernando, 2013]{Vecina:2013}
Vecina, M. L., \& Fernando, C. (2013). Volunteering and well‐being: is pleasure‐based rather than pressure‐based prosocial motivation that which is related to positive effects?. \textit{Journal of Applied Social Psychology, 43}(4), 870--878.

\bibitem[Wilson \& Post, 2013]{Wilson:2013}
Wilson, F., \& Post, J. (2013). Business models for people, planet (\& profits): Exploring the phenomena of social business, a market-based approach to social value creation. \textit{Small Business Economics, 40}(3), 715--737.

\bibitem[World Happiness Report, 2019]{WHR:2019}
World Happiness Report: Helliwell, J., Layard, R., \& Sachs, J. (2019). \textit{World Happiness Report 2019}, New York: Sustainable Development Solutions Network (\url{https://worldhappiness.report/ed/2019/}).



\end{thebibliography}

%
%

\end{document}